\documentclass[aps,showkeys,showpacs,onecolumn,nofootinbib,superscriptaddress]{revtex4}

\usepackage{graphicx}
\usepackage{amsfonts}
\usepackage{amsmath}
\usepackage{latexsym}
\usepackage{amssymb}
\usepackage[mathcal]{euscript}

\newcommand{\ket}[1]{| #1 \rangle}
\newcommand{\bra}[1]{\langle #1 |}

\newcommand{\Z}{\mathbb{Z}}

\newcommand{\pa}[1]{\left(#1\right)}

\newtheorem{Th}{Theorem}

\newtheorem{Lem}{Lemma}

\newtheorem{Def}{Definition}

\begin{document}

\title{
Intrinsically universal one-dimensional quantum cellular automata in two flavours
}

\author{Pablo Arrighi}
\email{pablo.arrighi@imag.fr} 

\author{Renan Fargetton}
\email{renan.fargetton@imag.fr} 

\author{Zizhu Wang}
\email{zizhu.wang@imag.fr}

\affiliation{University of Grenoble,
LIG, 46 avenue Felix Viallet, 38000 Grenoble, France}

\begin{abstract}
We give a one-dimensional quantum cellular automaton (QCA) capable of simulating all others. By this we mean that the initial configuration and the local transition rule of any one-dimensional QCA can be encoded within the initial configuration of the universal QCA. Several steps of the universal QCA will then correspond to one step of the simulated QCA. The simulation preserves the topology in the sense that each cell of the simulated QCA is encoded as a group of adjacent cells in the universal QCA. The encoding is linear and hence does not carry any of the cost of the computation. We do this in two flavours: a weak one which requires an infinite but periodic initial configuration and a strong one which needs only a finite initial configuration.
\end{abstract}



\maketitle

\section{Introduction}\label{introduction}

In this section we give a quick overview of QCA and explain why they matter. We then move on to explain why the notion of intrinsic simulation is so important in the CA community, we then list some related construction and provide an outline for the paper.

\subsection{Quantum cellular automata} 

\begin{figure}[h]
\centering
\includegraphics[scale=1.1, clip=true, trim=0cm 0cm 0cm 0cm]{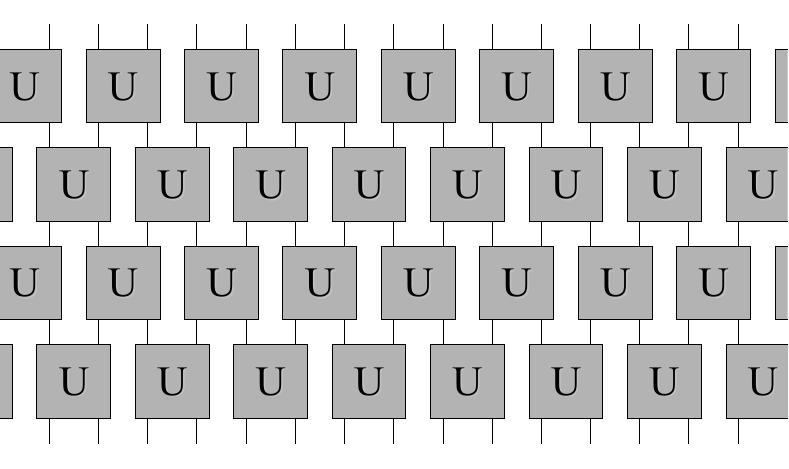}
\caption{\label{structure} Partitioned one-dimensional QCA with scattering unitary $U$. Each line represents a quantum system, in this case a whole cell. Each square represents a scattering unitary $U$ which gets applied upon two cells. Time flows upwards.}
\end{figure}

One-dimensional cellular automata (CA) consists of a line of cells, each of which may take one in a finite number of possible states. These evolve in discrete time steps according to a local rule, applied synchronously and homogeneously across space.  Because they are a physics-like model of computation it seems very natural to study their quantum extensions. The flourishing research in quantum information and quantum computer science provides us with appropriate context for doing so, both in terms of the theoretical framework and potential applications. Indeed this field has already brought to light a number of theoretical results about one-dimensional quantum cellular automata (QCA). For instance Schumacher and Werner \cite{Schumacher} have proved that any local, translation-invariant unitary evolution over a line of finite-dimensional quantum systems takes the form of a pattern of more elementary unitary evolutions -- repeated across time and space. And right from the very birth of the field with Feynman's 1986 paper, it was hoped that QCA may prove an important path to realistic implementations of quantum computers \cite{Feynman} -- mainly because they eliminate the need for an external, classical control and hence the principal source of decoherence. Other possible aims include providing models of distributed quantum computation, providing bridges between computer science notions and modern theoretical physics, or understanding the dynamics of some quantum physical system in discrete spacetime, i.e. from an idealized viewpoint. As we shall see in section \ref{definitions} of the paper, any QCA can be put into the form given by Figure \ref{structure}. For our purpose, we will take this particular space and time tiling of a scattering unitary $U$ as the actual definition of QCA.

\subsection{Intrinsic universality} 

\begin{figure}[h]
\centering
\includegraphics[scale=1.1, clip=true, trim=0cm 0cm 0cm 0cm]{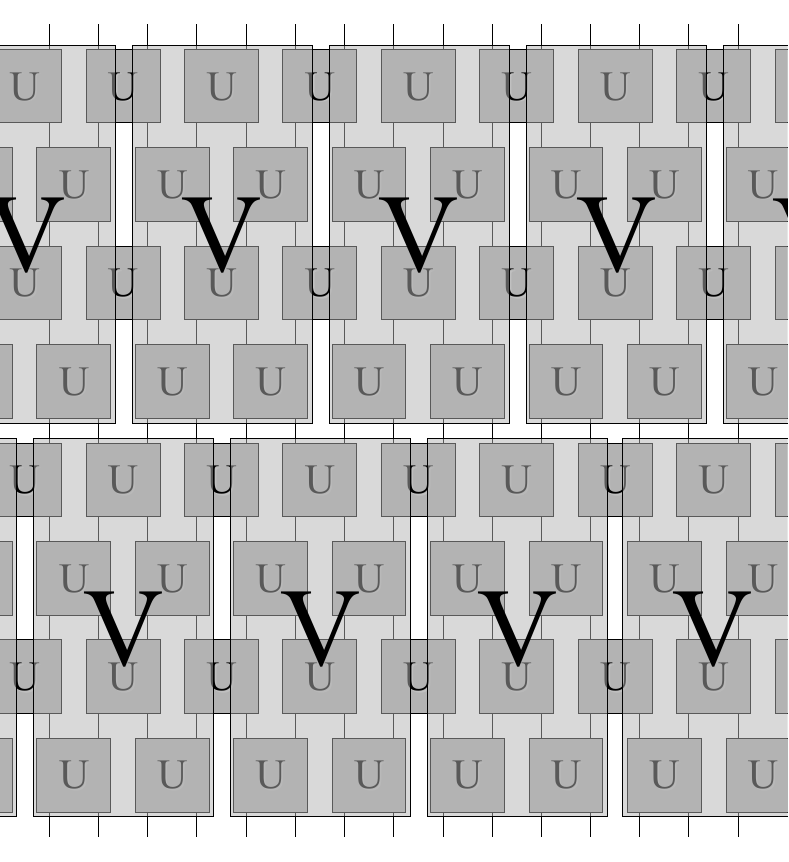}
\caption{Intrinsic simulation of a QCA by another.\label{UsimV}
(The QCA defined by $U$ simulates the QCA defined by $V$. In this case we need two cells of the $U$-QCA in order to encode one cell of the $V$-QCA, and we need to run the $U$-QCA for four time steps in order to simulate one time step of the $V$-QCA. More generally the challenge is to come up with an initial configuration of the $U$-QCA so that it behaves just as the $V$-QCA with respect to the encoded initial configuration, after some fixed number of time steps. Clearly such an encoding will have to hold the configuration of the $V$-QCA as well as some way of describing the scattering unitary $V$.)}
\end{figure}

\emph{The most popular cellular automaton is Conway's `Game of Life', a two-dimensional CA which has been proven to be universal for computation -- in the sense that any algorithm can be encoded within its initial state and then be run by the cellular automaton's evolution.} This was accomplished by simulating any Turing Machine (TM) within the automaton, and since Turing Machines have long been regarded as pretty much the best definition of `what an algorithm is' in classical computer science, this could have meant the end of the story to many people. Yet researchers in CA have always been looking for more than just running any algorithm, seeking to run distributed algorithms in a distributed manner, model some other phenomena together with their spatial structure, or make use of the spatial parallelism which is inherent to the model -- as these are the features which are modelled by CA and not by TM. And hence they have had to come back \cite{Albert} to the original meaning of the word `universal', namely the ability for one instance of a computational model to be able to simulate all other instances of the very same computational model. Nowadays there is an impressive number results about intrinsically universal CA as reviewed for instance in \cite{Delorme, Ollinger1} -- i.e. results on cellular automata capable of simulating all others efficiently and directly. (Incidentally of course they also simulate those CA which are capable of simulating the TM.) Or to put things differently, most of the CA community now seems to consider that a good notion of simulation is one which preserves the topology and the parallelism of the simulated CA, in some simple and explicit fashion. In the same manner, studying QCA rather than QTM for instance means we bother about the spatial structure of things, whether for the purpose of describing a quantum protocol, modelling a quantum physical phenomena, or again taking into account the spatial parallelism inherent to the model. Hence we argue that the kind of universality we are looking for is in fact stronger than the ability to simulate any quantum Turing Machine. We seek an intrinsically universal QCA, i.e. a QCA which can simulate all others efficiently and directly. A good intuition of the notion of intrinsic simulation is given by the diagram in Figure \ref{UsimV}, but formal definitions follow in section \ref{definitions}. In particular, we will distinguish two notions of intrinsic universality depending upon whether we allow preparation of an infinite periodic initial configuration or insist to have only finite configurations.

\subsection{Related results}

In the realm of classical computing Durand-Lose 
\cite{Durand}, has described an intrinsically universal one-dimensional reversible cellular automaton. Our construction will turn out to be a little simpler and cannot be substituted for this previous one, because reversible circuit universality requires at least one $3$-bit gate if done without the help of quantum mechanics. In the realm of quantum computing Shepherd, Franz and Werner \cite{Shepherd} have defined a class of QCA where the scattering unitary $V_i$ changes at each step $i$ (CCQCA). Via this construct they have built a QCA of cell-dimension $12$ which is universal in the circuit-sense. Universality in the circuit-sense had already been achieved by Van Dam \cite{VanDam} and Raussendorf \cite{Raussendorf} -- the latter uses a two-dimensional QCA but has this inspiring idea of programs crossing the data, with computation occurring in the interaction. Watrous \cite{Watrous} has proved that QCA are universal in the sense of Quantum Turing Machines. To our knowledge there is no previous work on intrinsically universal quantum cellular automata.

\subsection{Plan}

In section \ref{definitions} the reader is provided with the necessary theoretical background on QCA -- and the notion of intrinsic simulation is transposed to this theory. As in the classical case intrinsic simulation comes in two flavours, with one stronger than the other. In section \ref{weak} we construct a particular QCA, which we show is intrinsically universal, in the first sense. In section \ref{strong} we augment this QCA, and show that this results in a particular QCA which is intrinsically universal, in the second stronger sense. We conclude in section \ref{conclusion}.

\section{Theoretical background} \label{definitions}

\subsection{One-dimensional QCA}\label{subsecdef}

\noindent We now recall the fundamental definitions and properties of one-dimensional QCA. In what follows $\Sigma$ is a fixed finite set of symbols (i.e. `the  alphabet', describing the possible basic states each cell may 
take) and $q$ is a symbol such that $q\notin\Sigma$, which will be known as `the quiescent symbol', which represents an empty cells. We write $q+\Sigma=\{q\}\cup\Sigma$ for short.
\begin{Def}\textbf{(Finite configurations)} A \emph{(finite) configuration} $c$ over $q+\Sigma$ is a function $c:
\Z \longrightarrow q+\Sigma$, with $i\longmapsto
c(i)=c_i$, such that there exists a (possibly empty) finite
interval $I$ verifying $i\in I\Rightarrow c_i\in q+\Sigma$
and $i\notin I\Rightarrow c_i=q$. The smallest such interval $I$ is called interval domain of $c$, and is denoted $\textrm{idom}(c)$. The set of all finite configurations over $q+\Sigma$ will be denoted $\mathcal{C}_{q+\Sigma}$, whilst the set of all finite configurations having interval domain included in a finite interval $J$ will be denoted $\mathcal{C}_{q+\Sigma}^J$.
\end{Def}
\noindent Whilst configurations hold the basic states of an  entire line of cells, and hence denote the possible basic states of the entire QCA, the global state of a QCA may  well turn out to be a superposition of these. The following definition works because $\mathcal{C}_{q+\Sigma}$ is a countably  infinite set.\\
\begin{Def}\textbf{(Superpositions of configurations)}\label{superp} 
Let $\mathcal{H}_{\mathcal{C}_{q+\Sigma}}$ be the Hilbert space of configurations, defined as follows. To each finite configuration $c$ is associated a unit vector $\ket{c}$, such that the family $\pa{\ket{c}}_{c\in\mathcal{C}_{q+\Sigma}}$ is an orthonormal basis of $\mathcal{H}_{\mathcal{C}_{q+\Sigma}}$. A \emph{superposition of 
configurations} is then a unit vector in $\mathcal{H}_{\mathcal{C}_{q+\Sigma}}$. We also denote by  $\mathcal{H}_{\mathcal{C}_{q+\Sigma}^J}$ the subspace of $\mathcal{H}_{\mathcal{C}_{q+\Sigma}}$ spanned by the configurations in $\mathcal{C}_{q+\Sigma}^J$.
\end{Def}
\noindent Note that this space of QCA configurations is the same one as in \cite{Arrighi1,Arrighi2,Durr1,Durr2,Watrous}. It is isomorphic to the cyclic one considered in \cite{Meyer1}, but fundamentally different from the finite, bounded periodic space of \cite{VanDam} and the infinite setting of \cite{Schumacher}. The infinite setting of \cite{Schumacher} is slightly more permissive, but would force us to abandon the traditional setting of Hilbert spaces and move towards $C^*-$algebras. We choose not to do so out of simplicity, but our results apply to that context also since QCA have the same form in both contexts, as explained in \cite{Arrighi2}.
\begin{Def}\textbf{(Unitarity)}\label{unitarity} A linear operator $G:\mathcal{H}_{\mathcal{C}_{q+\Sigma}}\longrightarrow\mathcal{H}_{\mathcal{C}_{q+\Sigma}}$ is \emph{unitary} if and only if $\{G\ket{c}\,|\,c\in\mathcal{C}_{q+\Sigma}\}$ is an orthonormal basis of $\mathcal{H}_{\mathcal{C}_{q+\Sigma}}.$
\end{Def}
\begin{Def}\textbf{(Shift-invariance)}\label{shift-invariance} 
Consider the shift operation which takes configuration \\
$c=\ldots c_{i-1}c_ic_{i+1}\ldots$ to $c'=\ldots c'_{i-1}c'_ic'_{i+1}\ldots$ where for all $i$ $c'_i=c_{i+1}$. Let $\sigma:\mathcal{H}_{\mathcal{C}_{q+\Sigma}}\longrightarrow\mathcal{H}_{\mathcal{C}_{q+\Sigma}}$ be its linear extension to superpositions of configurations. A linear operator $G:\mathcal{H}_{\mathcal{C}_{q+\Sigma}}\longrightarrow\mathcal{H}_{\mathcal{C}_{q+\Sigma}}$ is said to be 
\emph{shift invariant} if and only if $G\sigma=\sigma G$.
\end{Def}
\begin{Def}\textbf{(Causality)}\label{locality} 
A linear operator $G:\mathcal{H}_{\mathcal{C}_{q+\Sigma}}\longrightarrow\mathcal{H}_{\mathcal{C}_{q+\Sigma}}$ is said to be 
\emph{causal} with with radius $\frac{1}{2}$ if and only if for any $\rho,\rho'$ two states over $\mathcal{H}_{\mathcal{C}_f}$, and for any $i\in\Z$, we have
\begin{equation}
\rho|_{i,i+1}=\rho'|_{i,i+1}\quad \Rightarrow G\rho G^{\dagger}|_i=G\rho'G^{\dagger}|_i. \label{loceq}
\end{equation}
\end{Def}
Here we used the notation $\rho|_J$ to mean restriction of $\rho$ to the region $J$ in the sense of the partial trace, and $G^{\dagger}$ is the Hermitian adjoint of $G$. In the classical case, the definition would be that the letter to be read in some given cell $i$ at time $t+1$ depends only on the state of the cells $i$ and $i+1$ at time $t$. Transposed to a quantum setting, we get the above definition: to know the state of cell number $i$, we only need to know the states of cells $i$ and $i+1$ before the evolution.\\

\noindent We are now ready to give the formal definition of one-dimensional quantum cellular automata.
\begin{Def}\textbf{(QCA)}\label{lca} 
A one-dimensional quantum cellular automaton (QCA) is an operator\\ $G:\mathcal{H}_{\mathcal{C}_{q+\Sigma}}\longrightarrow\mathcal{H}_{\mathcal{C}_{q+\Sigma}}$
which is unitary, shift-invariant and causal.
\end{Def}
This is clearly the natural axiomatic quantization of the notion of cellular automata. It was first given in \cite{Arrighi2} but stems from equivalent definition in the literature, phrased in terms of homomorphism of a $C^*$-algebra \cite{Schumacher}. There are other definitions in \cite{Arrighi1, Delgado, Durr1, Durr2, Meyer1, VanDam,Watrous} which are not axiomatic, in the sense that they all make particular assumptions about the form of the local action of $G$, and $G$ is then defined as a composition of these actions. That of \cite{Delgado} turns out to be almost equivalent to ours \cite{Arrighi2, Schumacher} in the end, as explained in \cite{Arrighi3}.

\subsection{Intrinsic simulation of one-dimensional QCA} \label{subsecsim}

The notions of intrinsic simulation of one CA by another arises with Banks in \cite{Banks}, but is not formalized until Albert and Culik \cite{Albert}. Even then, this apparently simple concept gives rise to two different  competing definitions, which we may call intrinsic simulations and strong intrinsic simulations respectively. Both are nicely explained in \cite{Delorme}, but subsequent works tend to focus on weak intrinsic simulations, as is nicely reviewed in \cite{Ollinger1, Ollinger2}. Hence let us start with an explanation of the more modern notion of intrinsic simulation.\\

\noindent The basic intuition in order to say that $G'$ simulates $G$ is that we can translate the content of each cell of $G$ into cells of $G'$, run $G'$, and then reverse the translation -- and that this three steps process will be equivalent to just running $G$. But first we must make it clear what we mean by `translate'. This translation should be simple (the idea is that the real cost of the computation is carried over by $G'$), it should preserve the topology (the idea is that each cell of $G$ is encoded into cells of $G'$ in a way which preserves whom neighbours whom), and it should be faithful (the idea is that no information should be lost in translation). This latter requirement translates into a precise notion in quantum theory, which is that of being an isometry, i.e. an inner product preserving evolution with $Enc^\dagger Enc=\mathbb{I}$. This same requirement also coincides with the fact that we would like this translation to be a physical process, i.e. that an actual translating machine can actually be built in theory. This is because quantum mechanics limits all physical evolutions to be isometries. With these observations in mind we reach the following definitions.

\begin{Def}\textbf{(Isometric coding)}\label{isomcode} 
Consider $q+\Sigma$ and $q''+\Sigma''$ two alphabets with distinguished quiescent states $q$ and $q''$, and such that $|q+\Sigma|\leq|q''+\Sigma''|$. Consider $\mathcal{H}_{q+\Sigma}$ and $\mathcal{H}_{q''+\Sigma''}$ the Hilbert spaces having these alphabets as their basis, and $\mathcal{H}_{\mathcal{C}_{q+\Sigma}}$, $\mathcal{H}_{\mathcal{C}_{q''+\Sigma''}}$ the Hilbert spaces of finite configurations over these alphabets.\\
Let $E$ be isometric linear map from $\mathcal{H}_{q+\Sigma}$ to $\mathcal{H}_{q''+\Sigma''}$ which preserves quiescence, i.e. such that $E\ket{q}=\ket{q''}$. It trivially extends into an isometric linear map $Enc=(\bigotimes_\mathbb{Z} E)$ from $\mathcal{H}_{\mathcal{C}_{q+\Sigma}}$ into $\mathcal{H}_{\mathcal{C}_{q''+\Sigma''}}$, which we refer to as an isometric encoding.\\
Let $D$ be isometric linear map from $\mathcal{H}_{q''+\Sigma''}$ to $\mathcal{H}_{q+\Sigma}\otimes\mathcal{H}_{q''+\Sigma''}$ which also preserves quiescence, in the sense that $D\ket{q''}=\ket{q}\otimes\ket{q''}$. It trivially extends into an isometric linear map $Dec=(\bigotimes_\mathbb{Z} D)$ from $\mathcal{H}_{\mathcal{C}_{q''+\Sigma''}}$ into $\mathcal{H}_{\mathcal{C}_{q+\Sigma}}\otimes\mathcal{H}_{\mathcal{C}_{q''+\Sigma''}}$, which we refer to as an isometric decoding.\\
The isometries $E$ and $D$ define an isometric coding if the following condition is satisfied:\\
$$\forall \ket{\psi}\in \mathcal{H}_{\mathcal{C}_{q+\Sigma}},\,\exists \ket{\phi}\in \mathcal{H}_{\mathcal{C}_{q''+\Sigma''}}\quad/\quad\ket{\psi}\otimes\ket{\phi}=Dec\pa{Enc \ket{\psi}}.$$
\end{Def}
(The understanding here is that $Dec$ is morally inverse function of $Enc$, but we may leave out some garbage $\ket{\phi}$ in the way.)

\begin{Def}\textbf{(Direct simulation)}\label{directsim}
Consider $q+\Sigma$ and $q''+\Sigma''$ two alphabets with distinguished quiescent states $q$ and $q''$, and two QCA $G$ and $G''$ over these alphabets. We say that $G''$ directly simulates $G$, if and only if there exists an isometric coding such that
$$\forall i\in\mathbb{N},\,\forall \ket{\psi}\in \mathcal{H}_{\mathcal{C}_{q+\Sigma}},\,\exists \ket{\phi}\in \mathcal{H}_{\mathcal{C}_{q''+\Sigma''}}\quad/\quad (G^i\ket{\psi})\otimes\ket{\phi}=Dec \pa{{G''}^i\pa{Enc \ket{\psi}}}.$$
\end{Def}

\noindent Unfortunately this is not quite enough for intrinsic simulation. Often we want to say that $G'$ simulates $G$ even though the translation:\\
- takes one cell of $G$ into several, not just one cell of $G'$;\\
- hence the quiescent symbol $q$ becomes a quiescent word $q'$;\\ 
- $G'$ needs be run $t$ times instead of just once.\\
All of these changes are made formal via the notion of grouping, as in the following definitions.

\begin{Def}\textbf{(Grouping)}\label{packmap} 
Let $G'$ be a QCA over alphabet $q'+\Sigma'$. Let $s$ and $t$ be two integers, $q''$ a word in $(q'+\Sigma')^s$, and $\Sigma''=\Sigma'^s \backslash \{q'\}$. Consider the iterate global evolution $G'^t$ up to a grouping of each $s$ adjacent cells into one supercell. If this operator can be considered to be a QCA $G''$ over $q''+\Sigma''$, then we say that $G''$ is an $(s,t,q')$-grouping of $G'$.
\end{Def}

\begin{Def}\textbf{(Intrinsic simulation)}\label{intsim}
Consider $q+\Sigma$ and $q'+\Sigma'$ two alphabets with distinguished quiescent states $q$ and $q'$, and two QCA $G$ and $G'$ over these alphabets. We say that $G'$ intrinsically simulates $G$ if and only if there exists $G''$ some grouping of $G'$ such that $G''$ directly simulates $G$.
\end{Def}

\begin{figure}
\centering
\includegraphics[scale=1.1, clip=true, trim=0cm 0cm 0cm 0cm]{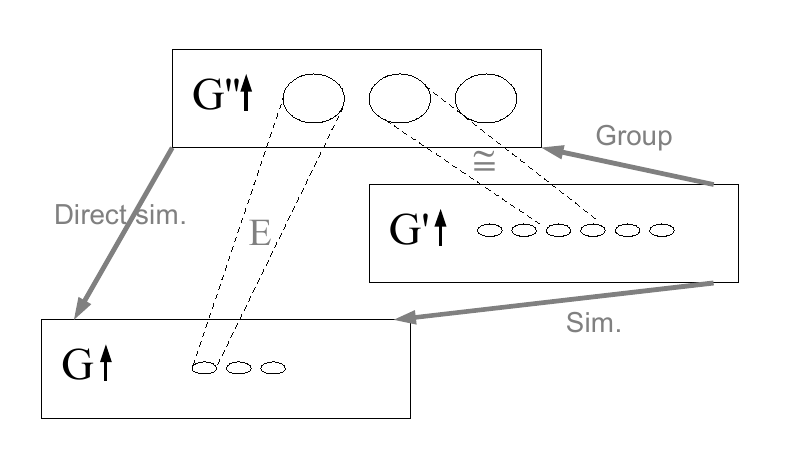}
\caption{The notion of intrinsic simulation made formal.\label{IntrinsicSim}}
\end{figure}

In simple words $G'$ intrinsically simulates $G$ if and only if there exists some isometry $E$ which translates each cell of $G$ into $s$ cells of $G'$, such that if we then run $G'^t$ and translate back, the whole process is just equivalent to just a run of $G$. This understanding is captured by Fig. \ref{IntrinsicSim}.
Notice, however, that $q$ the quiescent symbol of $G$  gets encoded into words $\ket{q'}=E\ket{q}$, which in general may not be the same as $\ket{q'^s}$, i.e. $s$ quiescent cells of $G'$. Hence with this notion of weak intrinsic simulation, we are indirectly assuming that the initial state of simulating QCA $G'$ could be prepared in a non-finite configuration, i.e. one which does not end and begin with only $q'$ symbols, but repeated $q'$ words instead. Formally this is not a problem, since $G'$ the $(s,t,q')$-grouping of $G'$ remains a valid QCA with quiescent symbol $q'$. Yet, depending on the application, one may wonder whether this notion of intrinsic simulation is the appropriate notion. For instance, if the implementation of $G'$ cannot be fed with $q'$ words left and right as the computation unravels, and if we do not know when the computation is supposed to stop, then this notion of intrinsic simulation may fail. These considerations will become clearer in the beginning of Section \ref{strong}, once illustrated with an example. Nevertheless they are exactly the ones which motivated the notion of strong intrinsic simulation.\\

\begin{Def}\textbf{(Strong isometric coding)}\label{strongisomcode}
Consider $q+\Sigma$ and $q''+\Sigma''$ two alphabets with distinguished quiescent states $q$ and $q''$, and such that $|q+\Sigma|\leq|q''+\Sigma''|$. Consider $\mathcal{H}_{q+\Sigma}$ and $\mathcal{H}_{q''+\Sigma''}$ the Hilbert spaces having these alphabets as their basis, and $\mathcal{H}_{\mathcal{C}_{q+\Sigma}}$, $\mathcal{H}_{\mathcal{C}_{q''+\Sigma''}}$ the Hilbert spaces of finite configurations over these alphabets.\\
Let $E, I$ be isometric linear maps from $\mathcal{H}_{q+\Sigma}$ to $\mathcal{H}_{q''+\Sigma''}$, where $I$ preserves quiescence, i.e. is such that $I\ket{q}=\ket{q''}$. Let $L, R$ be isometric linear maps from $\mathcal{H}^{\otimes r}_{q+\Sigma}$ to $\mathcal{H}^{\otimes r}_{q''+\Sigma''}$. Given any finite interval $J$, these can be combined into an isometric linear map $Enc^J=(I\bigotimes L\otimes \bigotimes_J E\otimes R \bigotimes I)$ from $\mathcal{H}_{\mathcal{C}_{q+\Sigma}}$ into $\mathcal{H}_{\mathcal{C}_{q''+\Sigma''}}$, which we refer to as the isometric encoding for interval $J$.\\
Let $D$ be isometric linear map from $\mathcal{H}_{q''+\Sigma''}$ to $\mathcal{H}_{q+\Sigma}\otimes\mathcal{H}_{q''+\Sigma''}$, in the sense that $D\ket{q''}=\ket{q}\otimes\ket{q''}$. It trivially extends into an isometric linear map $Dec=(\bigotimes_\mathbb{Z} D)$ from $\mathcal{H}_{\mathcal{C}_{q''+\Sigma''}}$ into $\mathcal{H}_{\mathcal{C}_{q+\Sigma}}\otimes\mathcal{H}_{\mathcal{C}_{q''+\Sigma''}}$, which we refer to as a strong isometric decoding.\\
The isometries $L,E,R$ and $D$ define a strong isometric coding if the following condition is satisfied:\\
$$\forall \ket{\psi}\in \mathcal{H}_{\mathcal{C}^J_{q+\Sigma}},\,\exists \ket{\phi}\in \mathcal{H}_{\mathcal{C}_{q''+\Sigma''}}\quad/\quad\ket{\psi}\otimes\ket{\phi}=Dec\pa{Enc^J \ket{\psi}}.$$
\end{Def}

In this scheme we now allow ourselves to know in advance whether the non-quiescent cells are all within the region $J$. Over the $J$ region we apply our encoding $\bigotimes E$ as usual. Over left and right quiescent tails we do not do much, since $\bigotimes I$ sends $\ket{\ldots qq\ldots}$ to $\ket{\ldots q''q''\ldots}$ anyway. In the direct surroundings of the $J$ region, we do something special, namely we apply $L$ on the left and $R$ on the right. Intuitively $L/R$ will be used to work at `encoding more left/right cells' as the configuration grows through the computation, leading to the following expected notion of strong direct simulation.

\begin{Def}\textbf{(Strong direct simulation)}\label{strgdirectsim} 
Consider $q+\Sigma$ and $q''+\Sigma''$ two alphabets with distinguished quiescent states $q$ and $q''$, and two QCA $G$ and $G''$ over these alphabets. We say that $G''$ strongly directly simulates $G$, if and only if there exists a strong isometric coding such that for all interval $J$ we have:
$$\forall i\in\mathbb{N},\,\forall \ket{\psi}\in \mathcal{H}_{\mathcal{C}^J_{q+\Sigma}},\,\exists \ket{\phi}\in \mathcal{H}_{\mathcal{C}_{q''+\Sigma''}}\quad/\quad (G^i\ket{\psi})\otimes\ket{\phi}=Dec \pa{{G''}^i\pa{Enc^J \ket{\psi}}}.$$
\end{Def}

\noindent Again this is not quite enough for intrinsic simulation, since we want to say that $G'$ simulates $G$ even though the translation takes one cell of $G$ into $s$ cells of $G'$, and one step of $G$ into $t$ steps of $G'$. But this time we want to make sure that the quiescent tails $\ket{\ldots qq \ldots}$ are taken into quiescent tails $\ket{\ldots q'q' \ldots}$. So we restrict the notion of grouping.

\begin{Def}\textbf{(Strong grouping)}\label{strongpackmap} 
Let $G'$ be a QCA over alphabet $q'+\Sigma'$. Let $s$ and $t$ be two integers,  and $\Sigma''=\Sigma'^s \backslash \{q'^s\}$. Consider the iterate global evolution $G'^t$ up to a grouping of each $s$ adjacent cells into one supercell. If this operator can be considered to be a QCA $G''$ over $q'^s+\Sigma''$, then we say that $G''$ is an $(s,t)$-strong-grouping of $G'$.
\end{Def}

\begin{Def}\textbf{(Strong simulation)}\label{strongdirectsim}
Consider $q+\Sigma$ and $q'+\Sigma'$ two alphabets with distinguished quiescent states $q$ and $q'$, and two QCA $G$ and $G'$ over these alphabets. We say that $G'$ strongly intrinsically simulates $G$ if and only if there exists $G''$ some strong grouping of $G'$ such that $G''$ strongly directly simulates $G$.
\end{Def}

Roughly we can say $G'$ strongly intrinsically simulates $G$ if and only if there exists some topology-preserving isometry $Enc$ which translates each finite configuration of $G$ into a finite configuration $G'$, such that if we then run $G'^t$and translate back, the whole process is just equivalent to just a run of $G$.

\subsection{The structure of one-dimensional QCA} \label{subsecstruc}

\begin{figure}
\centering
\includegraphics[scale=1.1, clip=true, trim=0cm 0cm 0cm 0cm]{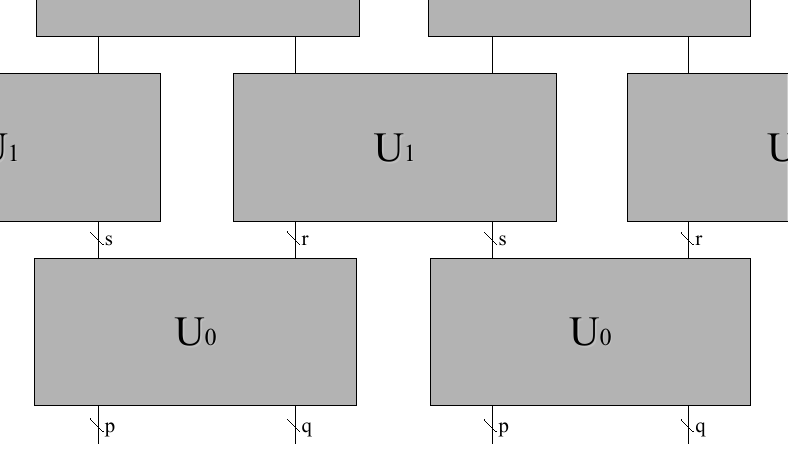}
\caption{Werner-style Margolus neighbourhood QCA \cite{Schumacher}.\label{werner} The elementary unitary evolutions $U_0$ and $U_1$ are alternated repeatedly as shown. The letters $p$, $q$, $r$, $s$ with $pq=rs$ denote the dimensions of the quantum systems.}
\end{figure}

\begin{figure}
\centering
\includegraphics[scale=1.1, clip=true, trim=0cm 0cm 0cm 0cm]{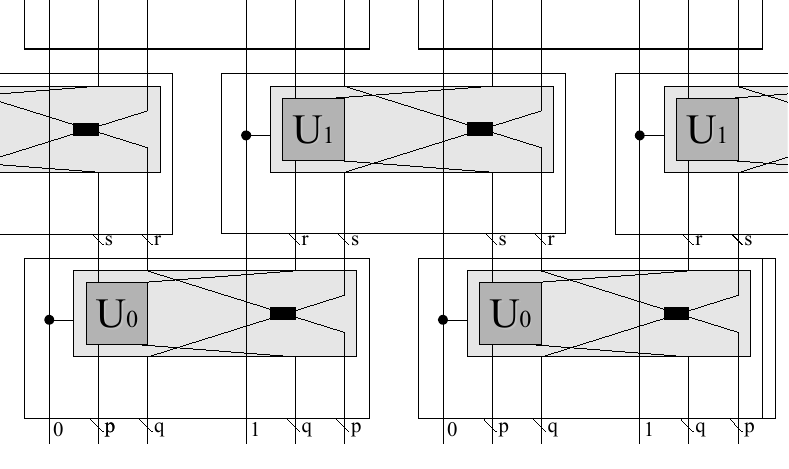}
\caption{Partitioned QCA simulating a Werner-style Margolus neighbourhood QCA.\label{Vsimwerner} The scattering unitary $U$ is a $(pq+1)^2\times(pq+1)^2$ matrix, essentially acting like either $U_0$ or $U_1$ on parts of the input according to the control qubit. The black box could be just any unitary, the lines upon which it acts are added just so that $U$ takes subsystems of an equal number of dimensions left and right.}
\end{figure}

\noindent As we have just seen one-dimensional QCA can be defined in quite general terms as a local, translation-invariant, global unitary evolutions $G$ acting over a line of finite dimensional quantum system. But contrary to its classical counterpart the definition does not yield an immediate way of constructing / enumerating all of the instances of this model. So if we just stick to that one definition QCA remain excessively abstract, hard-to-grasp mathematical object.\\ 
Fortunately Schumacher and Werner \cite{Schumacher}, and later Arrighi et al. \cite{Arrighi2} in the context of finite unbounded configurations, have proved that any such mathematical object is structured according to Figure \ref{werner}, i.e. that one-dimensional QCA admit a $2-$layered unitary block representation. These elementary unitary evolutions $U_0$ and $U_1$ are well-understood objects, i.e. they can be just any finite unitaries, as implemented for instance by some circuit of universal quantum gates for instance. Cheung and Perez-Delgado have proposed a definition of QCA directly in terms of local unitary evolutions, but thanks to this theorem the two are equivalent in one-dimension (the situation is similar in $n$-dimensions, see \cite{Arrighi3}).
\noindent This seems to be as good as it gets if we are not willing to modify the space upon which $G$ is acting. However the notion we are interested in is intrinsic universality, and in this context we are quite willing to interleave some extra cells for the purpose of simulating one QCA by another. And so it is not difficult to see that any such $U_0 U_1$-structured QCA can be straightforwardly made into a $U U$-structured QCA, as made explicit by Figure \ref{Vsimwerner}. 
Hence for the purpose of this paper, can really just restrict our attention to $U U$-structured QCA as in Figure \ref{structure}. These are also referred to as Partitioned QCA for instance in \cite{VanDam}, by analogy with classical Partitioned CA. 
\begin{Def}\textbf{(Partitioned QCA)}\label{pqca}
A partitioned  one-dimensional quantum cellular automaton (PQCA) over $\mathcal{H}_{\mathcal{C}_{q+\Sigma}}$ is defined by a unitary $U$ an operator $U:\mathcal{H}_{q+\Sigma}\otimes\mathcal{H}_{q+\Sigma}\longrightarrow\mathcal{H}_{q+\Sigma}\otimes\mathcal{H}_{q+\Sigma}$, such that $U\ket{qq}=\ket{qq}$, i.e. one which we take two cells into two cells and preserves quiescence. Let $G=(\bigotimes_{\mathbb{Z}} U)$ the operator over $\mathcal{H}_{\mathcal{C}_{q+\Sigma}}$. The induced global evolution is $G$ at odd time steps, and $\sigma G$ at even time steps, as in Fig. \ref{structure}.
\end{Def}
The defining elementary unitary evolution $U$ will be referred to as the scattering unitary, by analogy with quantum field theory. Hence we will talk about `$U$-defined QCA' in order to designate the PQCA having scattering unitary $U$. In this paper we demonstrate how a particular $U$-defined QCA can simulate any $V$-defined QCA, that is for any $V$. But again let us insist that this restriction to PQCA is without loss of generality, since we have just explained that any QCA can be put into the form of a PQCA. Notice that in Fig.~\ref{werner}, two vertical lines were denoting two subcells making up one cell of the QCA. But in the definition of PQCAs, we have referred to each of these subcells as just one cell. This was simply out of convenience, but one should keep in mind that one cell of a general QCA really corresponds to two cells of a PQCA.

\section{Intrinsic universality}\label{weak}

\subsection{What do we need for universality?}\label{conventions}

In subsection \ref{subsecdef} we have recalled the formal definition of One-dimensional QCA, and in subsection \ref{subsecstruc} we have recalled they have a simple circuit-like structure. Ultimately the picture one needs to have in mind in order to follow this section is that a QCA is just that of Fig. \ref{structure}.\\
In subsection \ref{subsecsim} we have provided a formal definition for the notion of intrinsic simulation. But again the picture one needs to have in mind in order to follow this section just that of Fig. \ref{UsimV}.\\
The purpose of this section is to find a particular $U$-defined QCA, and which is capable of intrinsically simulating any $V$-defined QCA, whatever the $V$. In order to describe that $U$-defined QCA we need to describe two things:\\
- How its cells are like (i.e. what are the vertical lines of Fig. \ref{structure} made of, what is the dimensionality of $U$?). By definition of QCA they are finite dimensional quantum systems of some fixed dimension $d$ of course, but for clarity we will decompose these into subsystems of dimension $d_i$, and we will give names to these subsystems according to their purpose (i.e. the vertical lines of Fig. \ref{structure} are `buses').\\
- How $U$ acts upon a pair of these cells, and more precisely upon the subsystems making up the pair of cells. We will say this informally, but we will also provide a formal circuit descriptions of $U$, and check that such a $U$ is indeed a unitary.\\
Before continuing with our detailed discussion of the QCA, it is useful for the sake of clarity to make precise some of the vocabulary and conventions that will be encountered in the rest of the paper. By a \emph{subsystem} we mean a constituent of a cell that serves a specific function in our QCA. A subsystem may take many different \emph{values}. Subsystems have names written in bold. The term \emph{signal} is used to refer to the value of a certain subsystem when it consistently travels between a cell and its left/right neighbour. In the space-time diagram of the QCA,  a signal looks like a `line' propagating through cells. Usually we take the name of a signal the name of the subsystem that takes those values.\\
This section will be organized as follows: first we give an intuitive idea about the mechanism we used to solve this problem, then we will explain different components of the QCA in detail, and finally we will see how the whole system fits together.

\subsection{Intuition}

\begin{figure}
\centering
\includegraphics[scale=1.1, clip=true, trim=0cm 0cm 0cm 0cm]{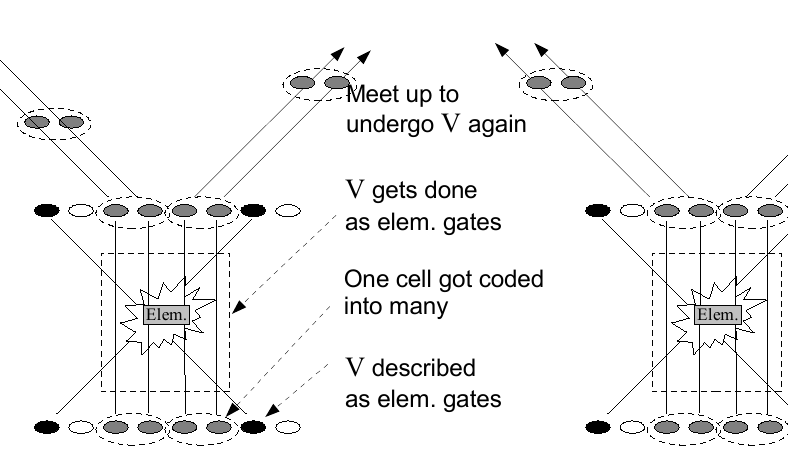}
\caption{Outline of the simulation scheme.\label{IntuitionWeak}}
\end{figure}

The cells of our universal, $U$-defined, QCA will have a subsystem called $\textbf{data}$ used for encoding one qubit of information about the state of a cell of the simulated, $V$-defined QCA. Hence one simulated cell (dotted oval in Fig. \ref{IntuitionWeak}) will in general be encoded as several adjacent simulating cells (small grey ovals in Fig. \ref{IntuitionWeak}). We also need to simulate the action of some arbitrary unitary $V$, and so the cells of our universal QCA will have a subsystem called $\textbf{program}$, holding some description of one of the elementary universal quantum gates that make up $V$. Hence $V$ (dotted box in Fig. \ref{IntuitionWeak}) will in general be encoded as several adjacent simulating cells (small black ovals in Fig. \ref{IntuitionWeak}). Because $V$ originally acts upon two cells, it is encoded in the surroundings of the two encoded cells.\\
During a first phase, as time unravels, the information held in the $\textbf{data}$ subsystem of the encoded cells remain stationary, and so we can think of them as some stationary data signals. Meanwhile the information held in the $\textbf{program}$ subsystems of the surroundings travels at lightspeed, and we can think of them as moving program signals. The program signals cross the data signals, leaving them unchanged, until they collide between one another. When that happens an elementary universal quantum gate is applied on the $\textbf{data}$ subsystems (grey box in Fig. \ref{IntuitionWeak}), thereby implementing $V$. Which elementary universal quantum gate gets applied depends on the value of the colliding program signals. Where the elementary universal quantum gate gets applied depends on where the collision takes place.\\
During a second phase, we need to `reload' this situation, with the added difficulty that $V$ gets applied in a shifted manner. Hence we need to arrange so that the left/right encoded cell travels left/right in order to meet with their right/left counterpart on the next site (travelling small grey ovals in Fig. \ref{IntuitionWeak}).\\
We will explain how to do this in three steps. First we draw a background pattern which allow us to synchronize this whole process. Second we let some data signals flow upon this background. Third we let the program signals move upon the background,  crossing and acting upon the data signals. Note that whenever we write $\ket{dgp}$ for the state of a cell, we mean that the subsystem $\textbf{data}$ is in state $\ket{d}$, subsystem $\textbf{program}$ is in state $\ket{g}$, subsystem $\textbf{mode}$ is in state $\ket{p}$.

\subsection{Ternary background pattern} \label{tern}

\begin{figure}
\centering
\includegraphics[scale=3, clip=true, angle=0, trim=2.97cm 1.30cm 2.98cm 1.80cm]{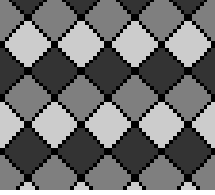}
\caption{Ternary background pattern. \label{ternary}
Each small coloured square now represents the state of a cell at time $t$. This is unlike in the previous figures, where boxes would represent a unitary evolution. At the bottom we have the initial configuration. Time flows upwards as usual. Each configuration is determined by the one below by pairing up the small squares of the configuration below and applying the scattering unitary $U$ to these pairs. The way they are paired up alternates in time: for odd steps cell $0$ is paired with cell $1$, cell $2$ with cell $3$, etc., whereas for even steps cell $1$ is paired up with cell $2$, cell $3$ with cell $4$, etc. The Light grey, Middle grey and Dark grey colours correspond to different values of the \textbf{mode} subsystem of the cells. They are separated by `Change colour' signals, represented in Black.
(With $\ket{000}$ in Light grey,  $\ket{001}$ in Middle grey, $\ket{002}$ in Dark grey, and $\ket{?1?}$ in Black.) }
\end{figure}

First we seek to draw the \emph{ternary background pattern} of Figure \ref{ternary}, where the large squares cycle through the three colours Light grey, Middle grey or Dark grey. The reason why this is useful will become clear in the next paragraph, for now it suffices to know that this ternary background pattern will help us synchronize the flow of the data signals as in Figure \ref{IntuitionWeak} and hence organize the computation as we add more interesting things to the initial configuration, i.e. the bottom of the diagram.\\
In order to achieve this ternary background pattern each cell must contain a $3$-dimensional system to code for those three different colours. This is really the purpose of subsystem \textbf{mode}: when the \textbf{mode} equals $0$, $1$ or $2$ the background colour is Light grey, Middle grey or Dark grey respectively. We must then place some signals at regular intervals, travelling at lightspeed and telling mode signals that they must change colour, and this is really role played by state $\ket{1}$ of subsystem \textbf{program}. \\
Let us show that the scattering unitary $U$ which is given in Figure \ref{Uascircuit} does the job of generating Figure \ref{ternary}. Observe Figure \ref{Uascircuit} and notice that the content of the \textbf{mode} and \textbf{program} subsystems is always propagated unchanged by the scattering unitary $U$, to the right/left if it comes from the left/right. Moreover observe Figure \ref{structure} and notice that at the next layer the content of \textbf{mode} and \textbf{program} will again come up from the left/right and hence be propagated again to the right/left. Hence whatever value is in the \textbf{mode} or in the \textbf{program} subsystem it just travels at maximal speed, right or left, depending only upon its position in the initial configuration. This is just what we mean by `a signal propagating at lightspeed'. In Subsection \ref{theqca} we provide all extra information needed about Figure \ref{Uascircuit} so that the behaviour of $U$ becomes fully-determined. We then state that $\ket{1}$ in the \textbf{program} subsystem is the control value required for the \textbf{+1 mod 3} to apply upon the \textbf{mode} subsystem. Hence the `Change colour' signals are indeed implemented by setting some cells to have their subsystem $\textbf{program}$ initialized at $\ket{1}$ as in Figure \ref{ternary}, and the `Change colour' signals indeed propagates at lightspeed, changing the value of the mode signals travelling at lightspeed in the opposite direction.\\
Notice that later, when we will set the \textbf{data} subsystem to non-$\ket{0}$ values in order to code for simulated cells, or use up the other possible values of the \textbf{program} subsystem in order to code for elementary universal gates to be applied upon the coded simulated cells, this ternary background pattern will remain unaffected. This is obvious from Figure \ref{Uascircuit} and the fact that $\ket{1}$ is the only value of \textbf{program} which triggers the \textbf{+1 mod 3} gate.\\
The required widths of the Light grey and Middle grey zones of the initial configuration vary depending upon the $V$-defined QCA we are seeking to simulate, in a way which we explain in Subsection \ref{hexa} and \ref{coll}, respectively. \\

\begin{figure}
\centering
\includegraphics[scale=3, clip=true, angle=0, trim=2.97cm 1.30cm 2.98cm 1.80cm]{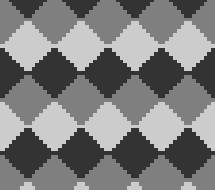}
\caption{Ternary background pattern. \label{ternary2}
This is the same as Figure \ref{ternary} except we no longer show the `Change colour signals'. (With $\ket{??0}$ in Light grey, $\ket{??1}$ in Middle grey, $\ket{??2}$ in Dark grey.)}
\end{figure}

\subsection{Hexagonal data signals flow} \label{hexa}

\begin{figure}
\centering
\includegraphics[scale=3, clip=true, angle=0, trim=2.97cm 1.30cm 2.98cm 1.80cm]{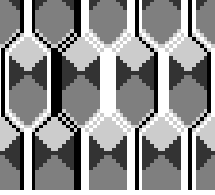}
\caption{Molding the hexagonal data flow pattern.\label{dataflow} The ternary background of Figure \ref{ternary2} is now used to tell the data signals when it is they should stay stationary, and when it is they should move to go and meet their counterparts.  We show the data signal as White for an encoded $\ket{0}$ qubit and Black for an encoded $\ket{1}$ qubit.
(With $\ket{0?0}$ in Light grey, $\ket{0?1}$ in Middle grey, $\ket{0?2}$ in Dark grey,
$\ket{1??}$ in White, $\ket{2??}$ in Black.)}
\end{figure}

Second we seek to draw the \emph{hexagonal data signals flow} of Figure \ref{dataflow}. That is we want to implement data signals, and would like that the data signals remain stationary for a while, and then separate into a left moving and a right moving bunch of data signals, only to eventually rejoin their left and right counterparts in order to flow straight in time again, as was explained in Figure \ref{IntuitionWeak}. As previously mentioned, the reason why we want to achieve this particular form of data signals flow is that it corresponds to the overall architecture of the QCA we are trying to simulate as in Figure \ref{UsimV}, with the interaction unitary $V$ taking its inputs as coming both from the left and the right, computing upon them, and spitting its outputs again both towards the left and the right for another run of $V$.\\
In order to achieve this each cell must contain another $3$-dimensional subsystem to code for a data signal. This is really the purpose of subsystem \textbf{data}: when the \textbf{data} equals $\ket{0}$, $\ket{1}$ or $\ket{2}$ the cell carries no data, an encoded $\ket{1}$ or an encoded $\ket{2}$ respectively.  We must then place the encoded data qubits in the Light grey coloured zones of the initial configuration, i.e. replacing the $\ket{000}$ cells by $\ket{100}$ cells in order to code for the presence of an encoded $\ket{0}$, and $\ket{000}$ by $\ket{200}$ in order to code for the presence of an encoded $\ket{1}$.\\
Let us show that the scattering unitary $U$ which is given in Figure \ref{Uascircuit} does the job of generating Figure \ref{dataflow}. The intuitive explanation will of course be that the Grey levels of the ternary background pattern are here to tell the data whether it can move or not, with the Middle and Dark grey forcing it to remain stationary, and the Light grey allowing it to move freely until they are gathered by a Middle grey funnel again. The more formal explanation relies on looking at the scattering unitary matrix $U$ which is given in Figure \ref{Uascircuit} in order to understand when the value of the left/right \textbf{data} subsystem is propagated to the right/left, and when it is just left sitting on the left/right.  This is what determines whether a data signal is stationary or moving at lightspeed. In order to have a complete answer to this question one must look at the definition of the $S$ gate in Figure \ref{Uascircuit}, as provided in Subsection \ref{theqca}. There we find that $S$ swaps the left and right \textbf{data} subsystems but only if one of them is $\ket{0}$ (i.e. a data signal moves right/left only if there is no data signal there) and if the values of both \textbf{mode} subsystems are $\ket{0}$ (i.e. the Light grey zones of the ternary background pattern). Initially the data signals are in a Light grey zones, but they are stuck upon another, so they cannot move. Not even the ones on the left and right ends can move -- due to the surrounding Middle grey zones. Clearly this stationary situation will be maintained until the ones on the left and right ends become surrounded by Light grey zones. During this period the data signals are freed two by two, with the one on the left end going to the left, and the one on the right end going to the right. Finally the first right moving data signal meets up with the first left moving data signal, and so they are stuck by one another for one step. But that one step is enough so that the second right/left moving data signal rejoins the first, so that the second one also gets stuck against the first, and the first does not bounce back. When all of the right/left moving signals have done so we are back to the original situation. Hence this ability of the ternary background pattern to `mold' the flow of data signals into the hexagonal shape of Figure \ref{dataflow}.\\
Notice that the required width of the Light grey zone of Figure \ref{ternary} has now become apparent. Since the purpose of this Light grey zone is to encode the state of two cells of the simulated QCA, it needs to be twice as big as the number of qubits that are needed to encode one cell of the simulated, $V$-defined QCA. 
\begin{figure}
\centering
\includegraphics[scale=3, clip=true, angle=0, trim=2.97cm 1.30cm 2.98cm 1.80cm]{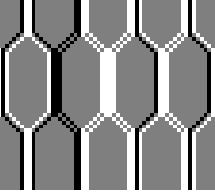}
\caption{Hexagonal data flow pattern.\label{dataflow2} This is the same as Figure \ref{dataflow} except we no longer show the ternary background.
(With $\ket{0??}$ in Grey, $\ket{1??}$ in White, $\ket{2??}$ in Black.)}
\end{figure}

\subsection{Collision gates}\label{coll}

\begin{figure}
\centering
\includegraphics[scale=5, clip=true, angle=0, trim=2.97cm 1.30cm 2.98cm 1.80cm]{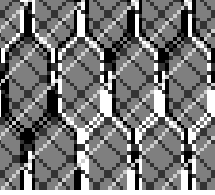}
\caption{Circuitry. \label{circuitry} The Middle grey initial zone of Figure \ref{dataflow2} has been modified in order to include two $\ket{3}$-valued program signals (here in Dark grey) and one $\ket{2}$-valued program signal (here in Light grey). This was done by changing two adjacent $\ket{001}$ to $\ket{031}$, and a $\ket{001}$ to $\ket{021}$. We can see that these program signals travel at lightspeed and sometimes collide. When they do so a two qubits elementary universal gate gets applied upon the data signals. The nature of the gate depends upon the values of the colliding program signals, whereas the position where the gate applies depends upon the place where the collision occurs. (Here whenever two $\ket{3}$-valued  program signals intersect a $cNot$ gate get applied to the data signals below them, whereas whenever a $\ket{2}$-valued program signal meets a $\ket{3}$-valued program signal a $\mathbb{I} \otimes Not$ gets applied. With $\ket{?3?}$ in Dark grey, $\ket{?2?}$ in Light grey, $\ket{20?}$ and $\ket{21?}$ in White, $\ket{20?}$ and $\ket{21?}$ in Black, and the rest in Grey.)}
\end{figure}

Third we seek simulate the scattering unitary $V$, as in Figure \ref{circuitry}. As was explained in Figure \ref{IntuitionWeak}, the key idea here is that during the time the data signals are stationary, they may be crossed by program signals incoming from both their left and their right, and sometimes these program signals even collide against one another upon the data signals. The value of the colliding program signals is what will specify \emph{what} should happen to the data signals, through a numbering of a set elementary universal quantum gates. The relative positions of the program signals is what will specify \emph{where} this should happen, by determining where they collide.\\
In order to achieve this we must change some of the $\ket{0}$ values of the $\textbf{program}$ subsystems of the Middle grey zones of the initial configuration, and allow them to take extra values $\ket{2}$ and $\ket{3}$. I.e. we will change some $\ket{001}$ cells into $\ket{021}$ or $\ket{031}$ cells and generate program signals carrying value $\ket{2}$ or $\ket{3}$, which will then travel at lightspeed, collide and so implement gates upon the data signals.\\
Let us show that the scattering unitary $U$ which is given in Figure \ref{Uascircuit} does the job of generating Figure \ref{circuitry}. We have already shown in Subsection \ref{tern} that the program signals travel at lightspeed, unaffected. The only thing we need to explain is what happens when they collide with one another. Again from Figure \ref{Uascircuit} and Subsection \ref{theqca} we have that whenever two program signals (\textbf{xprogram} and \textbf{yprogram} are both $\ket{2}$ or $\ket{3}$) cross each other upon some data signals (\textbf{xdata} and \textbf{ydata} non-$\ket{0}$) then some elementary quantum gate is applied upon \textbf{xdata}$\otimes$\textbf{ydata} as given in Table \ref{Mgate}. The mechanism is illustrated in Figure \ref{circuitry} but with essentially classical elementary gates -- so that we may draw their effect.\\ 
Notice that the required width of the Middle grey zone of Figure \ref{ternary} is starting to become apparent. Since the purpose of this Middle grey zone is to hold pairs of program signals, each pair coding for one of the elementary universal quantum gate implementing $V$, its size will depend will depend upon the number of elementary universal quantum gates of the circuit-description of $V$. However the exact size of those Middle grey zones will be determined in Subsection \ref{weakresults}.

\subsection{Fitting things together}\label{theqca}
So overall our universal, $U$-defined, QCA consists of a repeated application of one scattering unitary $U$ as in Figure \ref{structure}. Now follows the summarized description of the structure of the cells and the scattering unitary $U$.

\noindent \emph{Structure of the cells} 
In Figure \ref{structure} each vertical line do not represent just one qubit but a $36$-dimensional quantum system made of the subsystems described in the following table.
\begin{table}[!hbp]
\begin{normalsize}
	\centering
		\begin{tabular}{|c|c|p{10cm}|}
			\hline
				Name&Size&Function \\
				\hline
				\hline
				\textbf{data}&3& To hold one qubit of data of the QCA being simulated ($\ket{0}\equiv$Empty, $\ket{1}\equiv$Encoded $\ket{0}$, $\ket{2}\equiv$Encoded $\ket{1}$).\\
				\hline
				\textbf{program}&4& To code for what quantum gate should be applied to the data ($\ket{0}\equiv$Empty, $\ket{1}\equiv$Change colour, $\ket{2}/\ket{3}\equiv$cf. Table \ref{Mgate})\\
				\hline
				\textbf{mode}&3& To synchronize the flow of the data signals ($\ket{0}\equiv$Light grey $\ket{1}\equiv$Middle grey, $\ket{2}\equiv$Dark grey).\\
				\hline
		\end{tabular}
	\caption{Subsystems of the copy band}
	\label{bandsubsystems}
	\end{normalsize}
\end{table}

\begin{figure}
\centering
\includegraphics[scale=1.1, clip=true, trim=0cm 0cm 0cm 0cm]{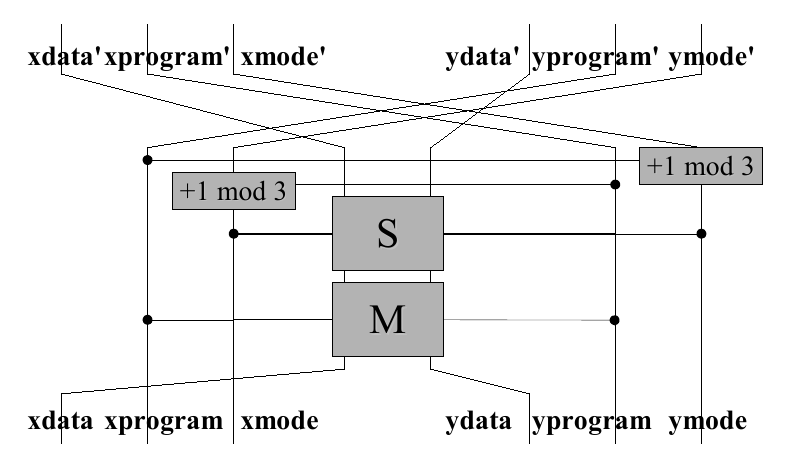}
\caption{The scattering unitary $U$ as a quantum circuit.\label{Uascircuit}
The three left/right lines represent the three subsystems making up the left/right cell. Each square represents an elementary quantum gate being applied of these subsystems, conditional to the value of the control systems designated by the horizontal lines. Time flows upwards. The whole thing represents the scattering unitary $U$, it takes two cells and yields back two cells.}
\end{figure}

\noindent \emph{Structure of the scattering unitary $U$.} 
In Figure \ref{structure} the scattering unitary $U$ takes two inputs \textbf{x} and \textbf{y}, each of which decomposes into three subsystems \textbf{xdata}, \textbf{xprogram}, \textbf{xmode} and \textbf{ydata}, \textbf{yprogram}, \textbf{yprogram} respectively, as mentioned. Therefore it could be given as a $36^2\times 36^2$  matrix of complex numbers, yet fortunately it decomposes as in Figure \ref{Uascircuit}.
In this diagram horizontal lines are control lines. Let us explain what we mean by control lines in general, as this will provide us with the main step of the proof of the unitarity of $U$. Say that a box labelled $B$ applies upon systems $S_1,\ldots, S_p$ whilst having an horizontal line crossing systems $T_1, \ldots, T_q$, then its effect is to apply the unitary evolution $B_i$ whenever the control system is in state $\ket{i}$ -- and linearly so. The following well-known lemma formalizes this construct.
\begin{Lem}[Unitarity of the control-construct]~\\
Let $S_1, \ldots, S_p$ and $T_1, \ldots, T_q$ be quantum systems of finite dimensions $|S_1|, \ldots, |S_p|, |T_1|, \ldots, |T_q|$. Let $\{B_i\}$ be a set of $|T_1|\cdots|T_q|$ unitary matrices of dimension $|S_1|\cdots|S_p|\times|S_1|\cdots|S_p|$. Then the evolution defined by linear extension of\\
$\ket{i}^{T_1\otimes\ldots\otimes T_q}\ket{\psi}^{S_1\otimes\ldots\otimes S_p}\mapsto \ket{i}^{T_1\otimes\ldots\otimes T_q}B_i\ket{\psi}^{S_1\otimes\ldots\otimes S_p}$
is itself unitary. Here the notation $\ket{\phi}^S$ means the quantum system $S$ in in state $\ket{\phi}$.
\end{Lem}
\textbf{Proof.} (This has appeared many times before in the literature but we reproduce it for completeness.) Since the evolution is a square matrix it suffices to check that it has orthonormal columns. \\ Let $\{\ket{i}^{T_1\otimes\ldots\otimes T_q}\}$ and $\{\ket{j}^{S_1\otimes\ldots\otimes S_p}\}$ denote the canonical orthonormal basis for $T_1\otimes\ldots\otimes T_q$ and $S_1\otimes\ldots\otimes S_p$ respectively. Then the inner product between column $ij$ and column $kl$ is given by\\
$(\bra{i}\otimes\bra{j}B_i^\dagger)(\ket{k}\otimes B_k\ket{l})=\delta_{ik}\bra{j}B_i^\dagger B_i\ket{l}=\delta_{ik}\delta_{jl}$.\hfill$\Box$\\
This being said we are now in a position to describe the three gates used, in terms of whatever canonical basis state their control systems may take. Of course the control systems do not actually have to be in a basis state, by linear extension and the above lemma. We shall see that the first two gates are used really just to move the encoded data qubits around, whereas the last gate actually does perform the computation.
\begin{itemize}
\item The \textbf{+1 mod 3} gate. If \textbf{xprogram}$\neq\ket{0}$ and \textbf{yprogram}$=\ket{4}$, then \textbf{xmode} is incremented by one modulo three. Else this is the identity. (Symmetrically so swapping the roles of \textbf{x} and \textbf{y}.) This step's role is to increment the modes, which in turn synchronizes the flow of the data signals. 
\item The \textbf{S} gate. If \textbf{xprogram}$=\ket{0}$, and \textbf{yprogram}$=\ket{0}$, then permute $\ket{01}$ with $\ket{10}$ and $\ket{02}$ with $\ket{20}$. Else this is the identity. This means we are swapping \textbf{xdata} and \textbf{ydata} only if one of them is empty and the modes are `White'. This step's role is to move the data when appropriate, in order to actually perform the flow of data.
\item The \textbf{M} gate. If \textbf{xdata}$\neq\ket{0}$, \textbf{ydata}$\neq\ket{0}$, \textbf{xprogram}$=\ket{2}$ and \textbf{yprogram}$=\ket{2}$, then the system \textbf{xdata}$\otimes$\textbf{ydata} undergoes an elementary quantum gate according to the state of the system \textbf{xmode}$\otimes$\textbf{ymode} as in Table \ref{Mgate}. This step's role is to apply a quantum gate upon two qubits of data, in order to perform the computation. 
\begin{table}
\begin{align*}
&\textbf{xprogram}\otimes\textbf{yprogram}\qquad   &\textbf{Action of M}\\
&\ket{22} &Swap &\\ 
&\ket{23} &\mathbb{I} \otimes H&\quad\textrm{Hadamard on the second qubit}\\
&\ket{32} &H \otimes \mathbb{I}&\quad\textrm{Hadamard on the first qubit}\\
&\ket{33} &cPhase&\\
&otherwise &\mathbb{I} \otimes \mathbb{I}
\end{align*}
where $cPhase$ stands for 
$\left(\begin{array}{cccc}
1 &0 &0 &0\\
0 &1 &0 &0\\
0 &0 &1 &0\\
0 &0 &0 &e^{i\pi/8}
\end{array}\right)$.\\
\caption{The \textbf{M} gate.\label{Mgate}}
\end{table}
\end{itemize}
In the next section we show how this all fits together to obtain the intrinsic universality. Beforehand however note that since we have described the evolution $U$ as a combination of smaller unitary matrices via tensors, composition and the control-construct, it is indeed unitary as required by Def. \ref{pqca}.

\subsection{Results}\label{weakresults}

\noindent \emph{Universal set of elementary quantum gates.} First let us show that we have a universal set of gates available in the QCA. The set of gates which the \text{M} gate is able to perform upon the data qubits has been chosen to be universal in the traditional sense, i.e. any finite dimensional unitary evolution $V$ can be approximated by tensors and compositions of these gates. We have not chosen the standard set ($cNot$, $H$, $Phase$) so as to preserve the \textbf{xy} symmetry of the unitary evolution $U$ and yet keep the dimension of \textbf{program} to a minimum, but it is easy to see that we can recover the standard set since 
\begin{align*}
cNot\ket{\psi}&=(\mathbb{I}\otimes H)(cPhase)^8(\mathbb{I}\otimes H)\ket{\psi}\\
\ket{1}\otimes Phase\ket{\psi}&=cPhase\ket{1}\otimes\ket{\psi}
\end{align*}
where the ancilla $\ket{1}$ can be brought over via applications of the $Swap$ gate.\\

\begin{figure}
\centering
\includegraphics[scale=1.1, clip=true, trim=0cm 0cm 0cm 0cm]{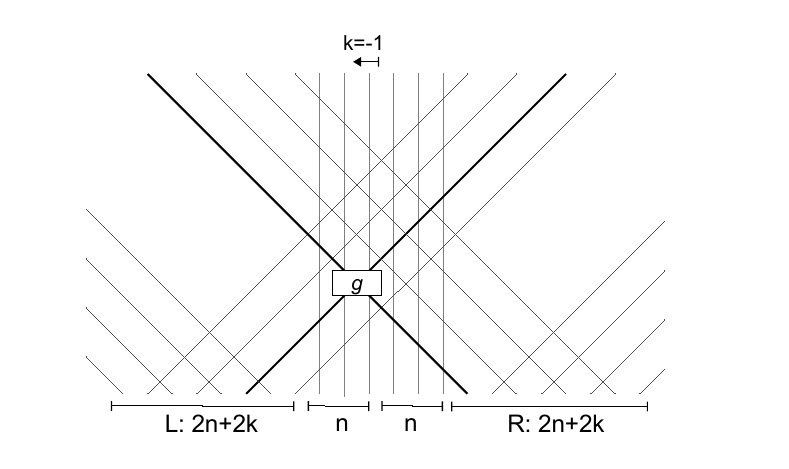}
\caption{Encoding a gate $g$.\label{encoding} Here the position of the gate relative to the center position is $k=-1$, meaning that the gate will be applied not on the center qubits but on the pair just left of it. And here the radius of the data signals bus is $n=3$, meaning that three qubits are needed to code for just one cell of the simulated QCA. The two black lines represent the right-moving and left-moving program signals that make up the encoding of the gate $g$. The fact that the right-moving program signal originates further apart from the stationary data signals than its left counterpart is what codes for the position $k=-1$. In the worst case scenario the position is $-n+1$. Clearly we must not send another left-moving signal until the worst case scenario right-moving signal has come out of the data signals, otherwise their collision may trigger the application of an unwanted gate. So each gate requires at least $2(n+k)$ cells on each side in order to be encoded. In the chosen encoding we systematically take $4n$ cells on each side in order to encode one gate.}
\end{figure}

\noindent \emph{Addressing each elementary quantum gate.} Second, let us show that we may combine our universal quantum gates in and arbitrary fashion so as to be able to implement any quantum circuit within the QCA. Consider some circuit $C$ made of $m$ elementary two qubit quantum gates $(g_i)_{i=0\ldots m-1}$ taken from our universal set of two-qubit gates ($Swap$, $\mathbb{I}\otimes H$, $H\otimes \mathbb{I}$, $cPhase$). Suppose that the circuit is $2n$ qubits wide, so that the positions $(p_i)_{i=0\ldots m-1}$ telling us where to apply those gates can be given as numbers in $-(n-1)\ldots (n-1)$ relative to the center.  For instance if $g_3=Swap$ and $p_3=0$, this means the fourth gate in the circuit consists in swapping the qubits at the center. Or say if $g_0=\mathbb{I}\otimes H$ and $p_0=n-1$, this means that the first gate of the circuit consists in applying a Hadamard to the rightmost qubit. Hence a position actually refers to a pair of qubits, position $-n+1$ being the leftmost pair of qubits, $-n+2$ the second leftmost pair etc. This is coherent with the fact that we have only two qubits gates in our chose universal set of quantum gates.\\
Then circuit as acting upon the initial state $\ket{\psi}$ can be encoded in our QCA via the state:
\begin{align*}
\ket{\Gamma(C,\psi)}&= 
(\bigotimes_{i=m-1}^{0}l_i)\otimes F\ket{\psi} \otimes (\bigotimes_{i=0}^{m-1}r_i).
\end{align*}
Here the $F\ket{\psi}$ region holds the qubits of the circuit, encoded within the cells of the $QCA$. The way to do this is as explained in Subsection \ref{hexa}: i.e. $\ket{0}/\ket{1}$ is encoded as $\ket{100}/\ket{200}$. (It is important that the presence of data $\ket{1??}/\ket{2??}$ be distinguished from the absence of data $\ket{0??}$.) Formally (with $\ket{\phi}^j$ the $j^{th}$ subsystem of $\ket{\phi}$):
\begin{align*}
F\ket{\psi}&=\bigotimes_{j=0}^{n-1}Inc\ket{\psi}^j\ket{00}\\
Inc\ket{0}&=\ket{1}\quad Inc\ket{1}=\ket{2}.
\end{align*}
The $(\bigotimes_{i=m-1}^{0}l_i)$ and $(\bigotimes_{i=0}^{m-1}r_i)$ hold the description of the circuit, encoded within the cells of the QCA. Say we want to perform the gate $g_i$ at position $p_i$ of these $2n$ encoded qubits. We will surround the qubits (and the previous gate descriptions) with a description of $(g_i, p_i)$, as in:
\begin{align*}
\textrm{with for $p_i\leq 0$, }\;
l_i&= \ket{011}^{\otimes 4n-1-2|p_i|}\otimes\ket{0l(g_i)1}\otimes\ket{001}^{\otimes 2|p_i|}\\
r_i&= \ket{0r(g_i)2}\otimes\ket{001}^{\otimes 4n-1}\\
\textrm{and for $p_i\geq 0$, }\;
l_i&= \ket{001}^{\otimes 4n-1}\otimes\ket{0l(g_i)1}\\
r_i&= \ket{001}^{\otimes 2|p_i|}\otimes\ket{0r(g_i)1}\otimes\ket{001}^{\otimes 4n-1-2|p_i|}\\
\textrm{moreover }\; l(Swap)&=r(Swap)=l(\mathbb{I}\otimes H)=2\\
l(cPhase)&=r(cPhase)=r(H\otimes \mathbb{I})=3
\end{align*}
This deserves some explanations. Say, as in the previous example, that we want for our fourth gate to have a $Swap$ happen at position $0$. We know from Table \ref{Mgate} that in order for $U$ to perform a $Swap$ upon its \textbf{data} subsystems, we need both \textbf{program} registers to be in state $\ket{2}$. So we need to have a program signal coming from the left and holding value $\ket{2}=\ket{l(Swap)}$, and the same from the right. Moreover we need them to originate equally far from the center so that they meet above the center, and we need to make sure that the next gate gets encoded far away enough so that its program signals do not intersect ours -- thereby inducing some unwanted operations. Finally the data signals need to be sandwiched by Middle grey mode signals so that they remain stationary, very much as explained in Subsection \ref{hexa}. Here $l_3=\ket{001}^{\otimes 2n-1}\otimes\ket{021}$ and $r_3=\ket{021}\otimes\ket{001}^{\otimes 2n-1}$ will do the job.  The rest is only a matter of placing the program signals in a adequate manner so that the gate is performed in the right position. Note that in the left/right encoded circuit description region, only one cell out of two is potentially used, because the other cells can only move in the opposite direction which will not go over our data, see Figure \ref{encoding}.
More convoluted encodings can be used to save up space and time, but only marginally. Also they seem to make the size of the encoding of a gate $g_i$ dependent upon its position $p_i$, which makes it more difficult to explain. What we have here does formalize the way to address each elementary quantum gate with respect to the data qubits in our scheme. It shows that we are able to simulate any such circuit $C$ by using up $8mn$ cells during $4mn$ steps. 

\noindent \emph{Intrinsic simulation of any partitioned one-dimensional QCA.} Third, let us show that this particular PQCA is intrinsically universal. In order to do this, consider a PQCA has scattering unitary $V$, and say we seek to approximate it just as illustrated in Figure \ref{UsimV}. In order to do so we must consider the quantum circuit $C(V)$ which approximates $V$ in terms of the quantum gates $(g_i)_{i=0\ldots m-1}$ acting at positions $(p_i)_{i=-n\ldots n-1}$. The approximated QCA as acting upon the initial state $\ket{\phi}$, can be encoded in our QCA via repetitions of the state $\ket{\Gamma(C,\phi,k)}$:
\begin{align*}
\ket{\Gamma(C,\phi,k)}&= 
\ket{011}\otimes w \otimes \ket{011}\otimes F\ket{\phi}^{2k,2k+1}\\
\textrm{with for $p_i\leq 0$, }\;
l_i&= \ket{011}^{\otimes 2n-1-|p_i|}\otimes\ket{0l(g_i)1}\otimes\ket{011}^{\otimes |p_i|}\\
r_i&= \ket{0r(g_i)1}\otimes\ket{011}^{\otimes 2n-1}\\
\textrm{and for $p_i\geq 0$, }\;
l_i&= \ket{011}^{\otimes 2n-1}\otimes\ket{0l(g_i)1}\\
r_i&= \ket{011}^{\otimes |p_i|}\otimes\ket{0r(g_i)1}\otimes\ket{011}^{\otimes 2n-1-|p_i|}\\
\textrm{moreover }\;
r&=\bigotimes_{i=0}^{m-1} r_i,\;l=\bigotimes_{i=m-1}^{0} l_i\\
w&=\bigotimes_k l^k \otimes r^k
\end{align*}
Here the scheme is exactly that of Fig. \ref{circuitry}. Compared to the previous coding here is what deserves explanations. The data is encoded exactly in the same way, namely each pair of cells $\ket{\phi}^{k,k+1}$ is encoded as $2n$ cells $F\ket{\phi}^{k,k+1}$. Then the description of the circuit $C$ is also encoded in the same way, in the surroundings of each encoding cells. There are a couple of differences however. Previously we were using only one out of two cells in the left/right encoded circuit description region, but now we can interleave the left and right encoded circuit descriptions to use them all, this is what $w$ does and the reason why the factor of $2$ was dropped in the $l_i/r_i$ descriptions. Also, we now need to handle mode and `Change colour' signals appropriately so as to obtain the ternary background pattern and the hexagonal data signals flow. By inspection of Fig. \ref{dataflow} we see that this is done by setting the left/right encoded circuit description regions to having subsystems \textbf{mode} at `Middle grey' and separating them from the encoded pair of cells with a `Change colour' signal (the $|011\rangle$ cells).\\
Again this is not quite optimal, but it does entail the following result.
\begin{Th}
There exists $G'$ a $U$-defined QCA which is intrinsically universal QCA in the following sense. Let $G$ be $V$-defined QCA such that $V$ can be expressed as a quantum circuit $C$ made of $m$ gates acting upon $2n$ qubits. Then $G'$ is able to intrinsically simulate $G$ with space expansion factor $s=4nm+2+2n$ and time expansion factor $t=(3/2)s$.
\end{Th}
\textbf{Proof.} 
Each cell of the $V$-defined QCA will be encoded into a string of cells of the $U$-defined QCA which we have described in this Section, according to the formula $\ket{\Gamma(C,\phi,k)}$ given in this Subsection (remember that that the two cells $2k$ and $2k+1$ of a PQCA are really just the $k$ cell of the corresponding QCA in the original definition, as we explained in Subsection \ref{subsecstruc}). This constitutes our isometric encoding $E$. The size of $\ket{\Gamma(C,\phi,k)}$ can be seen to be $4nm+2+2n$, which explains the value of the space expansion factor $s$. The ratio between the time expansion factor and the space expansion factor is deduced by inspection of Figure \ref{circuitry}.\\
\hfill $\Box$
Note that if the scattering unitary $V$ is only approximated with an error of $\epsilon=\max_{\ket{\psi}}||V\ket{\psi}-C(V)\ket{\psi}||$ by the quantum circuit $C(V)$, then this entails we are able to intrinsically approximate the evolution of $s$ cells over $t$ steps with an error bounded by $st\epsilon$ -- again using supercells of size $s=4nm+2+2n$ and a time expansion of factor $t=(3/2)s$. This is the general statement that errors in quantum circuits grow no more than proportional to time and space \cite{Nielsen}, which stems from the fact that if $||U-U'||\leq\epsilon$ then $||U^{\otimes s}-U'^{\otimes s}||\leq s\epsilon$ and $||U^t-U'^t||\leq t\epsilon$.

\section{Strong intrinsic universality} \label{strong}
The idea of tackling the problem of strong intrinsic universality was suggested to us by Torsten Franz, whom we would like to thank.
\subsection{Why do we need another flavour?}
\begin{figure}[!htbp]
\centering
\includegraphics[scale=1.75]{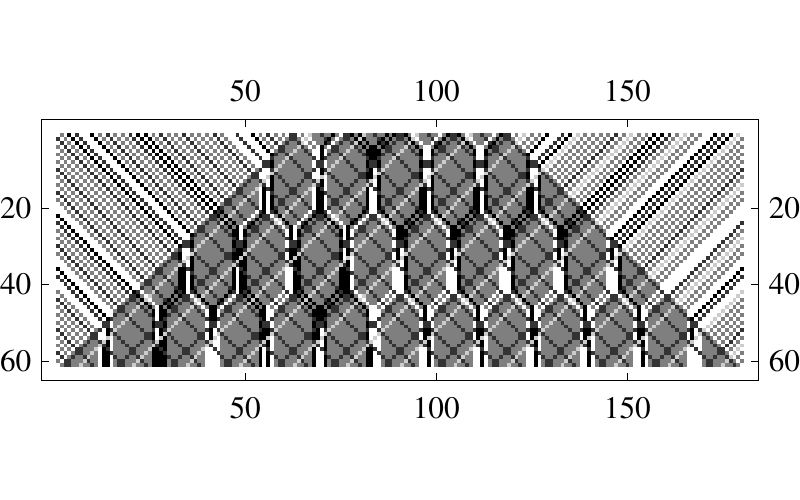}
\caption{The ternary background pattern will shrink as the computation goes on and the surrounding quiescent cells enter the computation region.}
\label{dispersingbg}
\end{figure}

The QCA explained in subsection ~\ref{theqca} has one problem: one needs to be able to prepare an infinite initial configuration somehow. If we do not do that, then the quiescent cells surrounding the background pattern will slowly begin to mix with the pattern, thus rendering it unusable. As can be seen in Fig. ~\ref{dispersingbg}, after certain amount of time, 
the whole ternary background pattern will be reduced to a single cell, and no more computation is possible. Thus, in order to perform an arbitrary long computation, we have to know the length of the computation beforehand, so that we can prepare an initial configuration 
of adequate width and protect the ternary background pattern long enough for the computation to finish. But in practice finding out how long the computation is going to last is a very difficult task, and may even be undecidable. 
In order to solve this problem, we need a second flavour to the notion of intrinsic universality, namely strong intrinsic universality, as was formally presented and defined in Subsection~\ref{subsecsim}. Hence we shall build a new QCA that has the ability to weave the ternary background pattern as the computation proceeds.
This section will be organized as follows: first we give an intuitive idea about the mechanisms we use to solve this problem, then we explain the different components of the QCA in detail, and finally we explain how the whole system fits together.

\subsection{Intuition}
In John Conway's `Game of Life' CA, there are objects called `guns' that can fire cells in different directions at different time intervals. Here we use a similar idea in order to fire the signals that make up the ternary background pattern. Essentially, we use what we call a `copy band', i.e. a structure made out of two walls 
and a synchronization signal bouncing between them, as in Fig.~\ref{copyband}. The distance between the walls controls the frequency of duplication, and the bouncer signal decides when to fire a piece of signal constituting the ternary background. Recall that this background was essentially made out of two signals: the mode (the \textbf{mode} subsystem is described in subsection ~\ref{theqca}), and the program (each program signal carries half of the information needed to code for a universal quantum gate, refer back to Table ~\ref{Mgate} to see how each gate is encoded as the state of two \textbf{program} subsystems). The left/right half of the coding of a universal quantum gate will hence be held by a left/right copy band. Since the fired mode and program signals travel at maximum speed, they appear on the space-time diagram of the QCA as having a 45/135-degree angle with respect to the horizontal line. So the two program signals each encoding one half of the program will eventually meet in the middle, inducing the application of a quantum gate. The program signals are being duplicated at a constant rate, thus creating the `squares' in Fig. ~\ref{ternary}. Part of the difficulty is to determine what level of grey should be sent (the value of the mode signal). This will be explained further down. The speed of the computation, manifested in Fig. ~\ref{ternary} as the size of the `squares', can be controlled by adjusting the distance between the walls and hence the frequency of the bounces.

\subsection{The copy band}
\begin{figure}[!htbp]
\begin{center}
\includegraphics[scale=0.6,clip=true]{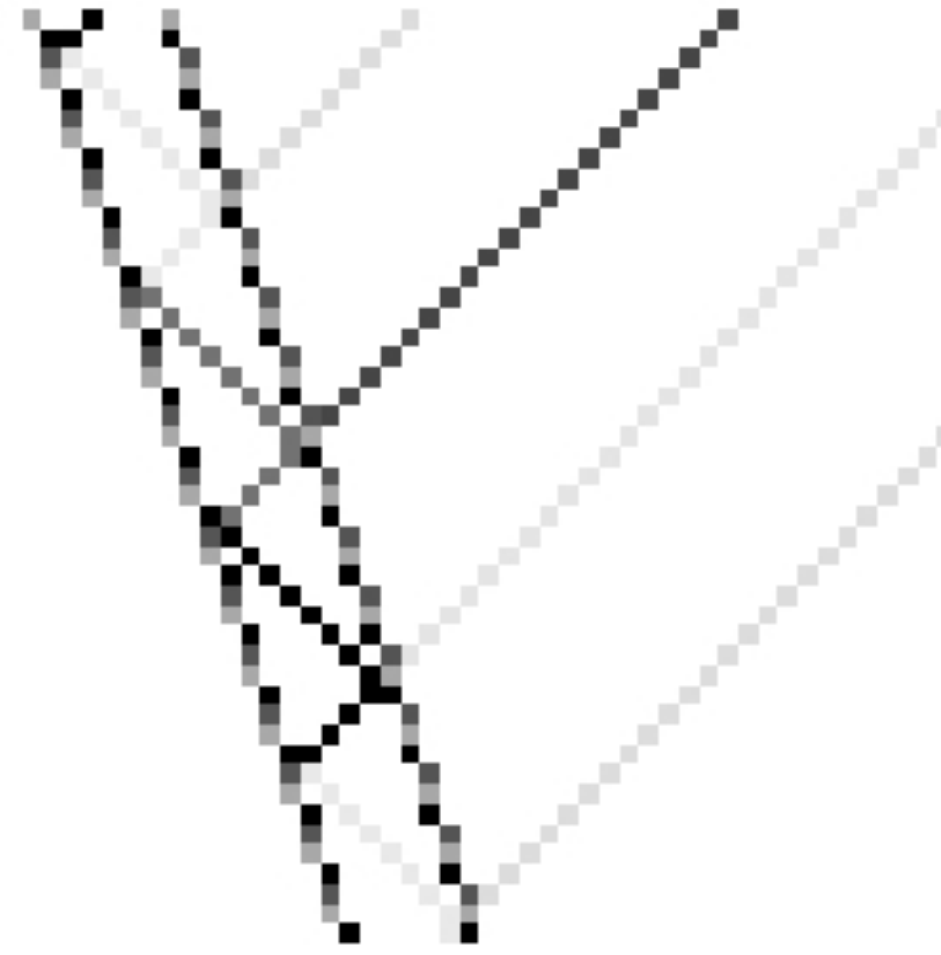}
\caption{A band. Both walls of the band contain a counter signal, and the left wall also contains a copy of the program signal which will be fired later in order to make up the ternary background pattern. The bouncer signal can be seen bouncing between the walls, with different grey level corresponding to different values of this signal. The duplicated program signal propagates to the right at light speed. Time flows upwards.}
\label{copyband}
\end{center}
\end{figure}

We continue with the convention described in Subsection~\ref{conventions}, i.e. the names of the subsystems will be given in bold (Examples of subsystems are the \textbf{mode} and \textbf{program} subsystems in Table~\ref{bandsubsystems2}), and names of the signals will not (An example of a signal is the bouncer signal in Fig.~\ref{copyband}).

\noindent A copy band is shown in Fig. ~\ref{copyband}. It is composed of a left wall, a right wall and the bouncer signal traveling in between. The subsystems used in order to implement the copy band are summarized in Table ~\ref{bandsubsystems2}. 

\begin{figure}[!htbp]
\begin{center}
\includegraphics[scale=0.6,clip=true]{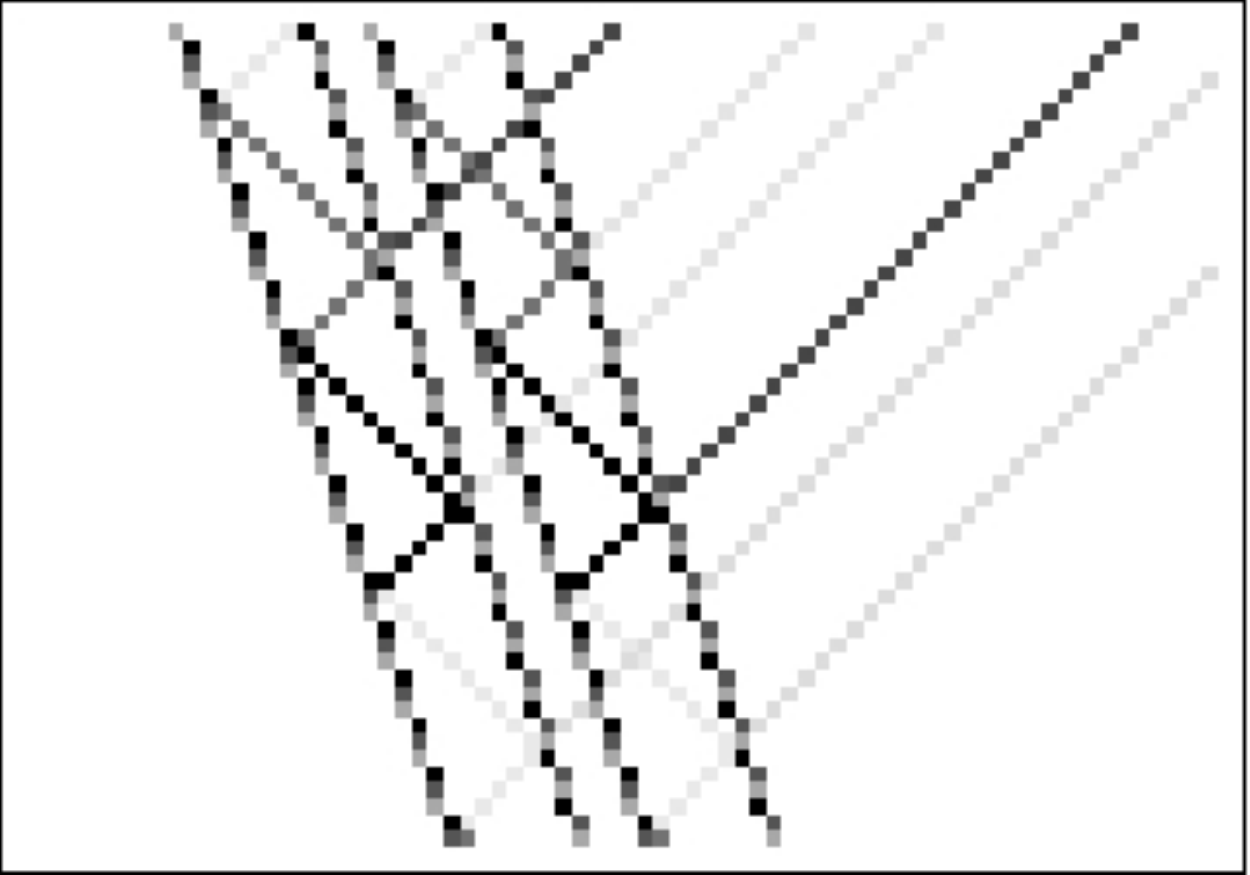}
\caption{Two copy bands placed side by side. The band will not alter a program signal that is already traveling at light speed, so the program signals emitted by the band to the left just pass through the band to the right. Time flows upwards.}
\label{twocopybands}
\end{center}
\end{figure}

\begin{table}[!hbp]
\begin{scriptsize}
	\centering
		\begin{tabular}{|c|c|p{10cm}|}
			\hline
				Name&Size&Function \\
				\hline
				\hline
				\textbf{counter}&4&The counter guarantees that the ternary background pattern grows at a constant rate. It increments by one at each step, and whenever it gets to 3, the copy band will start to shift left (or symmetrically, right ) to transform one quiescent cell into part of the ternary background pattern.\\
				\hline
				\textbf{wall}&3&Walls are what mark the border of a `copy band', separates it from the quiescent cells and the ternary background pattern. Cells that are not part of any copy band have value $|0\rangle$ (i.e. quiescent cells, the ternary background pattern and data cells). The walls that are closer to the half-line of quiescent cells have value $|1\rangle$, they are referred to from now on as `outer walls'. The right walls in the left half of the QCA have value $|2\rangle$, they are referred to as `inner walls' from now on.\\
				\hline
				\textbf{bouncer}&4&The bouncer is a signal that travels inside the \emph{copy band}, and changes its value whenever it hits a wall of value $|1\rangle$. Every time the bouncer bounces on the outer wall, a copy operation is performed on the \textbf{copy} subsystem. The resulting copy is put in the \textbf{program} subsystem, which travels at maximum speed to construct gates in the ternary background pattern. Together with the \textbf{history} subsystem, it enables calculation of the correct \textbf{mode} value for the fired ternary background signal.\\
				\hline
				\textbf{history}&3&To calculate the correct mode of the duplicated program bit, a `history' of how many mode-change signals have been encountered by this copy band is indispensable. Together with the \textbf{bouncer} subsystem, the correct mode can then be inferred. \\
				\hline
		\end{tabular}
	\caption{Subsystems of the copy band}
	\label{bandsubsystems2}
	\end{scriptsize}
\end{table}

\noindent The program signal that encodes half the information of a quantum gate is kept in the outer wall. This wall also stores the counter, which is used to slow down the speed of the walls by counting from $1$ to $3$ and then letting the walls move right/left only when the value of the counter signal is $3$, the $0$ meaning that there is no counter signal. The bouncer signal is placed next to this wall in the initial configuration. When the computation begins, the bouncer signal will travel towards the opposite wall at maximum speed. When it hits the outer wall, its value will change. The value of the bouncer signal is essentially another counter that cycles from $1$ to $3$, the $0$ meaning that there is no bouncer signal. Hence each change of value is just the sequence of operation $-1,\,(+1\mod{3}),\, +1$, applied only when the value differs from $0$. For simplicity we refer to this operation as $Inc_3$.
\noindent Whenever the bouncer signal hits the outer wall, the program/mode signals that constitute the background pattern will be fired. Namely, a copy of the value of the \textbf{copy} subsystem (cf. Table ~\ref{allsubsystems}) will be placed in the \textbf{program} subsystem, and the value of the \textbf{mode} subsystem will be set appropriately. The way this is done involves the value of the  \textbf{counter} and the \textbf{history} subsystems, and will be explained in detail in subsection ~\ref{modesync}.

\begin{figure}[!htbp]
\begin{center}
\includegraphics[scale=0.6]{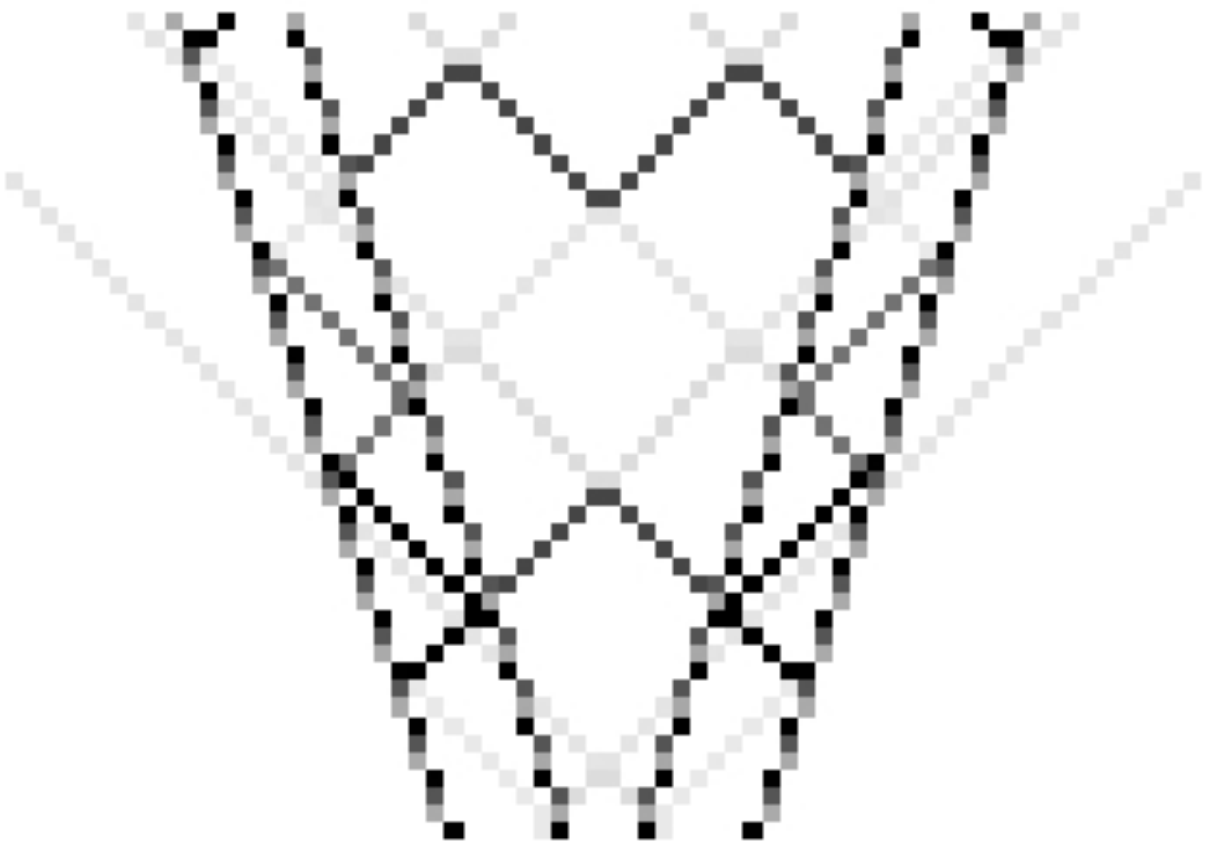}
\caption{Universal quantum gates encoded by two symmetrically placed copy bands. Gates are formed when two duplicated programs traveling in opposite directions intersect. Seven gates can be seen in the graph.}
\label{gates}
\end{center}
\end{figure}

\noindent Since each copy band only holds half the information needed to code a quantum gate in the ternary background pattern, another band is needed in order to reconstruct the gate. By exploiting the symmetry of the QCA depicted in subsection~\ref{theqca}, we use two symmetrically placed bands to `weave' one gate in the ternary background pattern, 
as shown in Fig.~\ref{gates}

\subsection{Mode synchronization}\label{modesync}

In subsection ~\ref{theqca}, all gates start in the same mode at time 0, and the mode cycles  each time a `change colour' signal is encountered. Since program signals start and travel at the same speed towards each other, it takes them an equal amount of time to meet in the middle, meeting
 the same number of `change colour' signals on the way, thus making mode synchronization automatic. In the QCA described in this section, however, gates start in different modes at different times, and although they still travel at the same speed 
 towards each other, they may very well meet different numbers of `change colour' signals on the way, so when two program signals meet, their mode may be out of sync, which may wrongly render this gate inactive/active. To address the problem, we used the information stored in the \textbf{history} subsystem and value of the \textbf{bouncer} subsystem in order to compute the correct value for this mode signal.

\noindent The bouncer signal decides when to fire the program signal and in which mode. It has three possible values, each corresponding to a value of the fired mode signal. The bouncer signal cycles through these values each time it `touches' the outer wall (i.e. wall of value $|1\rangle$), at this time the program stored in the outer wall will also be duplicated and fired as a signal, with the current value of the bouncer signal as its default mode, so to say.

\begin{figure}[!htbp]
\begin{center}
\includegraphics[scale=0.4]{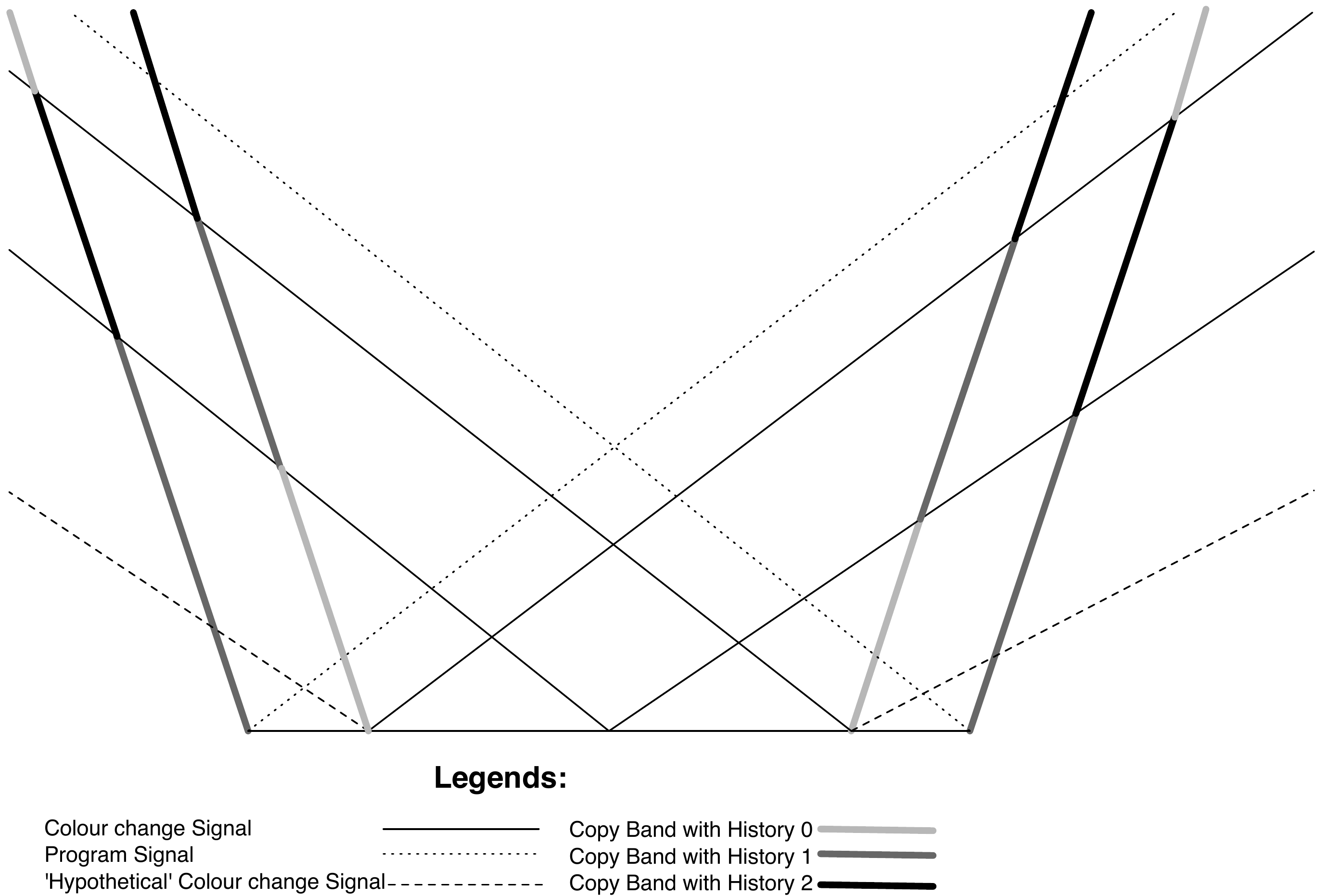}
\caption{Synchronization of the mode of the fired ternary background signal.  The initial value of the \textbf{history} subsystem should be the number of  `change colour' signals the copy band would have been encountered had the ternary background pattern been complete (i.e. had the `hypothetical' signals actually been sent). In the depicted situation, the value of the \textbf{history} subsystem of the outer copy bands would be $1$ because the outer bands would have encountered a `change colour' signal if the region occupied by the copy bands had been part of the initial configuration. 
}
\label{sync}
\end{center}
\end{figure}

\noindent The \textbf{history} subsystem stores the number of `change colour' signals that a program signal `would have encountered' if the signal had been sent at time 0. The effects of these `hypothetical' signals cannot be ignored, we must keep track of them modulo $3$. So actually the value of the fired mode will be that of the bouncer *plus* the offset kept by \textbf{history}. This mechanism is illustrated in Figure~\ref{sync}.
 
\subsection{Fitting things together}\label{strongfitting}

\subsubsection{Components and circuit overview}
Table~\ref{allsubsystems} summarizes the subsystems that make up a cell. In addition to the ones already mentioned in Table~\ref{bandsubsystems} and ~\ref{bandsubsystems2}, there is also the \textbf{copy} subsystem, which is used to store the value of the program signal to be fired. 
\begin{table}[!hbp]
\begin{scriptsize}
	\centering
		\begin{tabular}{|c|c|p{10cm}|}
			\hline
				Name&Size&Function \\
				\hline
				\hline
				\textbf{counter}&3&cf. Table~\ref{bandsubsystems2}.\\
				\hline
				\textbf{wall}&3&cf. Table~\ref{bandsubsystems2}.\\
				\hline
				\textbf{bouncer}&4&cf. Table~\ref{bandsubsystems2}.\\
				\hline
				\textbf{history}&3&cf. Table~\ref{bandsubsystems2}. \\
				\hline
				\textbf{data}&3&cf. Table~\ref{bandsubsystems}.\\
				\hline
				\textbf{copy}&4&The system that carries half the information needed to construct a gate. This information is permanently stored in a copy band, and will be duplicated when a ternary background signal needs be fired. It has the same size as the \textbf{program} subsystem of subsection~\ref{theqca}.\\
				\hline
				\textbf{program}&4&cf. Table~\ref{bandsubsystems}.\\
				\hline
				\textbf{mode}&3&cf. Table~\ref{bandsubsystems}.\\
				\hline
		\end{tabular}
	\caption{Subsystems of the QCA}
	\label{allsubsystems}
	\end{scriptsize}
\end{table}

\begin{figure}[!htbp]
\begin{center}
\includegraphics[scale=0.4]{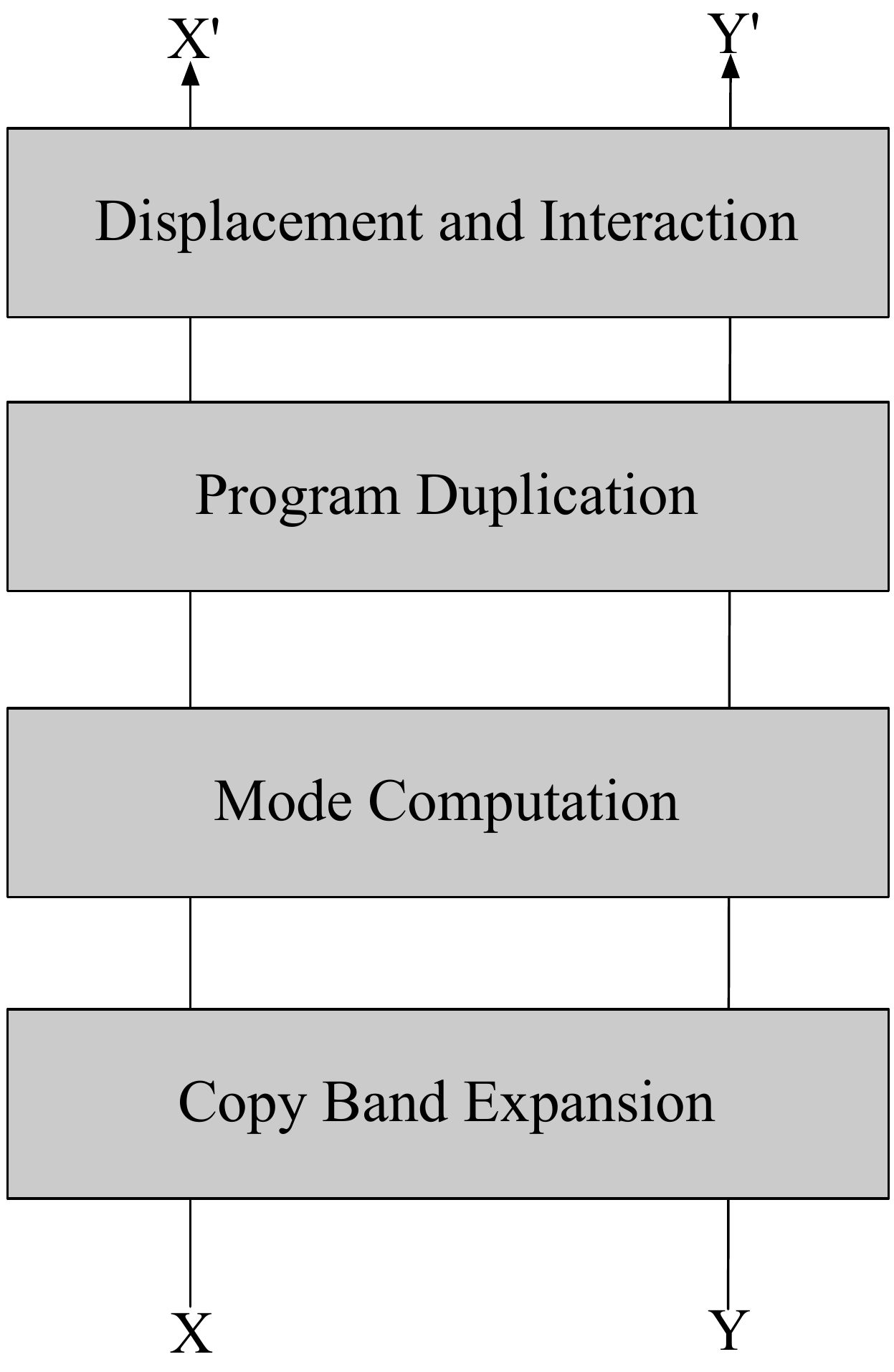}
\caption{The schematic layer diagram of the QCA's circuit. The \textbf{X} and \textbf{Y} lines denote the left and right cells undergoing the transformation. Time flows upwards.}
\label{layers}
\end{center}
\end{figure}

The circuit of the QCA has a similar structure to the circuit depicted in Fig.~\ref{Uascircuit}, albeit being more complex. To help visualize the circuit, it is divided into several different `layers', with each layer performing one or two basic functions. 
A schematic diagram of the different layers is given in Fig.~\ref{layers}. We will layout the circuit diagrams for the different layers and explain their functions in detail below. We follow the same conventions in the detailed diagrams as in Fig.~\ref{Uascircuit} (with horizontal lines representing control lines, and different subsystems represented by different vertical lines, time flowing upwards. The \textbf{x}s denote subsystems of the left cell, and the \textbf{y}s denote subsystems of the right cell.)
 
\subsubsection{Layers of the QCA's circuit}
\begin{itemize}
\item The Copy Band Expansion Layer (Fig.~\ref{expansion_layer}): This layer `expands' the ternary background two new cells at a time (one cell one the left, one cell on the right) by shifting the walls of the copy bands outwards, converting quiescent cells into part of the background pattern. Because the \textbf{copy} and the \textbf{history} subsystems are stored in the walls, they follow the expansion of \textbf{counter} and \textbf{wall} subsystems. The \textbf{counter} subsystem controls the swapping of \textbf{copy}, \textbf{history} and \textbf{wall} subsystems. That is, the swap gates are applied whenever the value of \textbf{xcounter} or \textbf{ycounter} is $|3\rangle$. The $SC$ gate performs the function of swapping the value of \textbf{xcounter} and \textbf{ycounter} whenever one of them is $|3\rangle$.  It is unitary because it is essentially just a permutation of the different base vectors of the subsystems upon which it acts. Its permutation table is given in Table~\ref{sc}. This layer also lets the bouncer signal travel and bounce, i.e. the values of \textbf{xbouncer} and \textbf{ybouncer} are swapped whenever the value of \textbf{xwall} and \textbf{ywall} are both $|0\rangle$ (i.e. the bouncer signal is traveling between the walls). Depending on the position of the copy band (i.e. the band is at the left or right side of the ternary background), only one $Inc_3$ gate will be applied. This gate is to cycle the value of the bouncer signal when the value of \textbf{xwall} or \textbf{ywall} is $|1\rangle$.
\begin{figure}[!htbp]
\begin{center}
\includegraphics[scale=0.5]{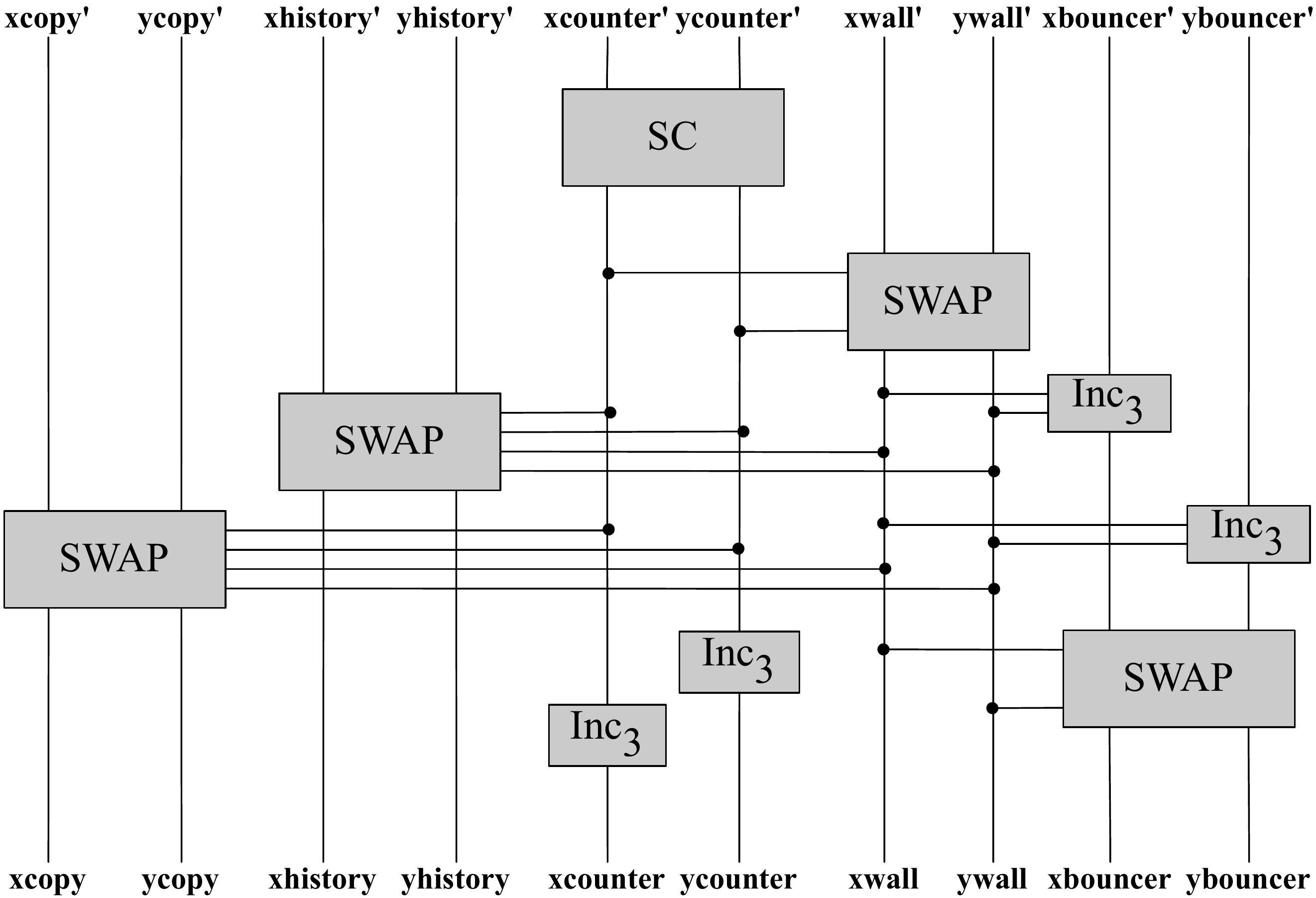}
\caption{The copy band expansion layer.}
\label{expansion_layer}
\end{center}
\end{figure}

\begin{table}[!hbp]
\begin{scriptsize}
	\centering
		\begin{tabular}{|c|c|p{10cm}|}
			\hline
				\multicolumn{2}{|c|}{The $SC$ gate}\\
				\hline
				\hline
				Input&Output\\
				\hline
				$|30\rangle$&$\langle03|$\\
				\hline
				$|03\rangle$&$\langle30|$\\
				\hline
				$|31\rangle$&$\langle13|$\\
				\hline
				$|13\rangle$&$\langle31|$\\
				\hline
				$|32\rangle$&$\langle23|$\\
				\hline
				$|23\rangle$&$\langle32|$\\
				\hline
				$|ij\rangle (i\neq3,j\neq3)$&$\langle ij| (i\neq3,j\neq3)$\\
				\hline
		\end{tabular}
	\caption{The $SC$ gate.}
	\label{sc}
	\end{scriptsize}
\end{table}

\item The Mode Computation Layer (Fig.~\ref{mode_layer}): this layer implements the rest of the synchronization mechanism discussed in subsection~\ref{modesync}. Here both the \textbf{counter} and the \textbf{history} subsystems must be used to determine the correct mode of the fired program signal. The \textbf{bouncer} subsystem contains the `base mode' of the program signal. We call it `base mode' because in order to compute the correct mode, we must `offset' the base mode by the value of the \textbf{history} subsystem. The function of the $CM$ gate is to `offset' the base mode by the value of the \textbf{history} subsystem, giving a new mode and leaving the \textbf{history} subsystem unchanged. To understand this one must know that even though the ternary background signal is fired at regular time intervals by the outer wall of a  copy band, its mode is only worked out when it crosses the inner wall and leaves out the copy band. Hence the $CM$ gate is only applied when the value of \textbf{xwall} is  $|0\rangle$, the value of \textbf{ywall} is $|2\rangle$, to make sure that the gate is applied at the inner wall, and when there is a non-empty \textbf{xprogram} (to make sure that a ternary background signal is crossing this inner wall). The $CM$ gate shown in the figure only acts on half of the copy bands, namely the copy bands that fire program signals to the \emph{right}. Because the QCA is symmetric, the $CM$ gate that acts on the other half can be obtained by replacing every $x$ in the figure with $y$. Another function played by this layer is the incrementation of the history. Recall that the value of \textbf{history} is incremented when the inner wall that holds the \textbf{history} subsystem meets a program signal with value $|4\rangle$ (cf. the \textbf{program} subsystem in Table~\ref{bandsubsystems}). But the inner wall may also meet the program signal fired by the outer wall of the same copy band. To differentiate these two program signals, we can use the bouncer signal: the program signal fired by the outer wall always travel with the bouncer signal, which is non-zero, until it acquires its mode and leaves the copy band. On the other hand, the program signal coming from the `outside' of the copy band does not have a bouncer signal traveling with it. Contrary to the $CM$ gate, the $+1\mod{3}$ gate shown in the figure acts on copy bands which fire program signals to the \emph{left}. The $+1\mod{3}$ gate is applied when the value of \textbf{xprogram} is $|4\rangle$, and the value of \textbf{xbouncer} is not zero (to make sure the `change colour' signal comes from the outside of the copy band), and the value of \textbf{ywall} is $|2\rangle$ (to make sure we are applying the gate to the \textbf{yhistory} subsystem in the inner wall).

\begin{figure}[!htbp]
\begin{center}
\includegraphics[scale=0.4]{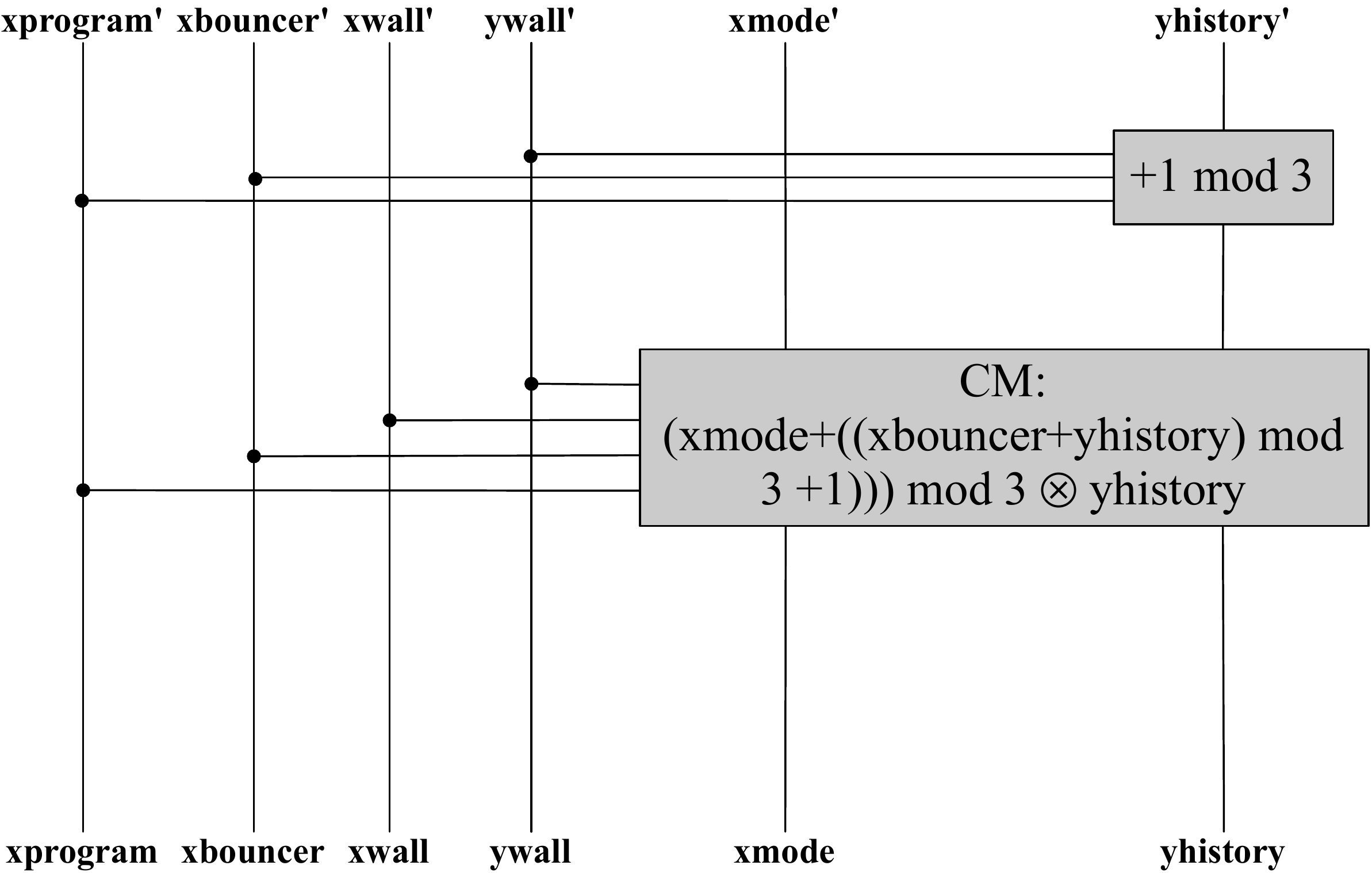}
\caption{Left half of the mode computation layer.}
\label{mode_layer}
\end{center}
\end{figure}

\item The Program Duplication Layer (Fig.~\ref{program_layer}): the function of the $CP$ gate in this layer is to change the value of \textbf{program} subsystem to the value stored in \textbf{copy} subsystem whenever a bouncer signal meets the outer wall that stores the \textbf{copy} subsystem. It is also responsible for incrementing the value of the bouncer signal.  This figure also shows only the left half of the actual diagram. The $CP$ gate is applied when the value of \textbf{xwall} is $|0\rangle$, the value of \textbf{ywall} is $|1\rangle$ (to make sure the gate is applied to the outer wall), and a non-zero \textbf{xbouncer} (to make sure that the bouncer signal has reached the outer wall).
\begin{figure}[!htbp]
\begin{center}
\includegraphics[scale=0.4]{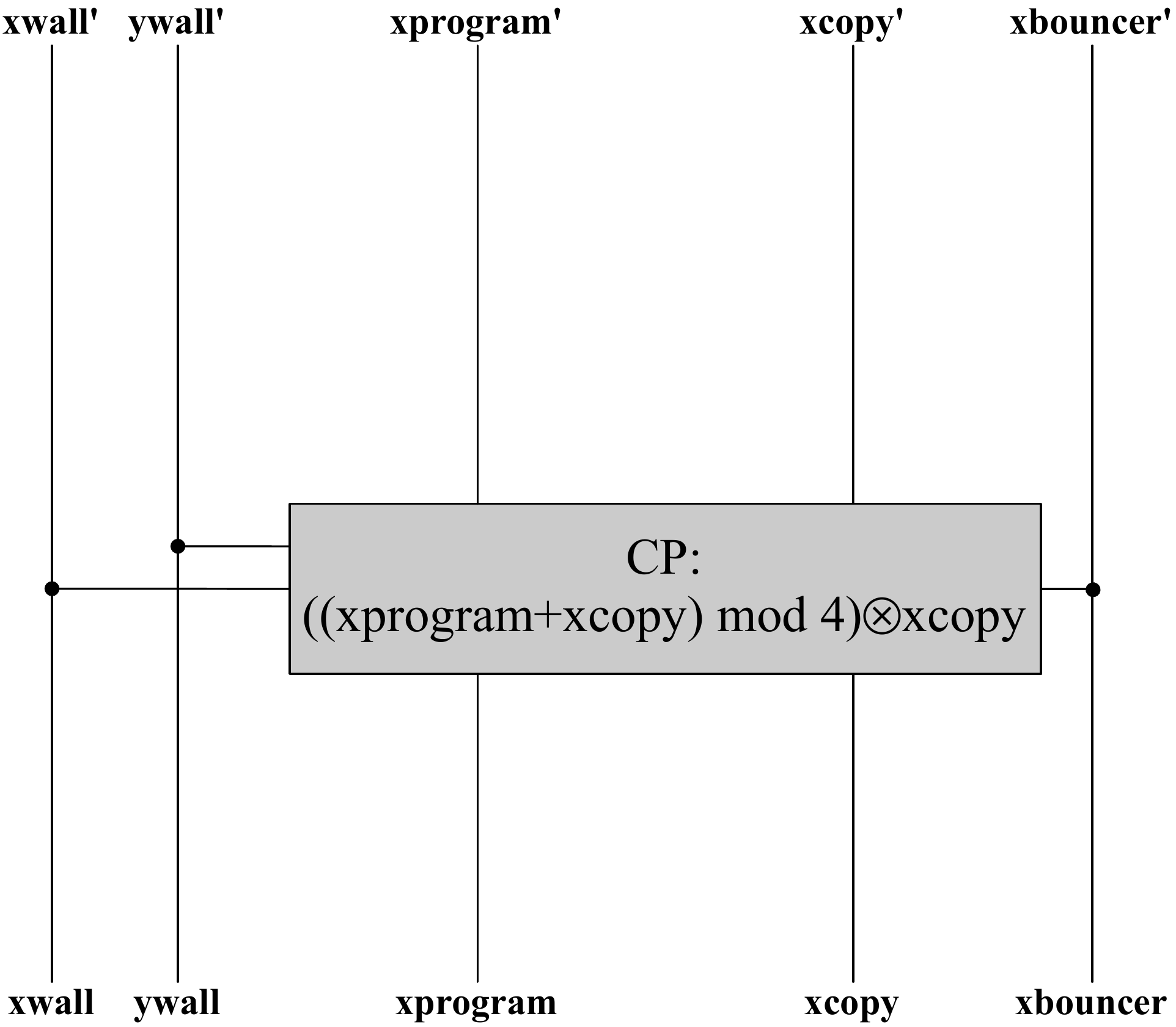}
\caption{Left half of the program duplication layer.}
\label{program_layer}
\end{center}
\end{figure}

\item The Displacement and Interaction Layer: this layer is identical to the scattering unitary $U$ of the weak intrinsic universal QCA described in subsection~\ref{theqca} (cf. Fig.~\ref{Uascircuit}). It only acts on the three subsystems given in Table~\ref{bandsubsystems} while leaving the other subsystems unchanged. It is responsible for moving the duplicated program and data around, and performing the computation by having the reconstructed gates act on the data.

\end{itemize}

\subsubsection{Initial state preparation}\label{initprep}
We now turn our attention towards the question of how to set up the initial state of out strong intrinsic QCA in order to simulate some $V$-defined QCA. This will be very much alike what we did in Subsection~\ref{weak}. But for the QCA described in this section we must also set up the copy bands, and special attention must be paid to the initialization of the mode synchronization mechanisms.

\begin{figure}[!htbp]
\begin{center}
\includegraphics[scale=1.2]{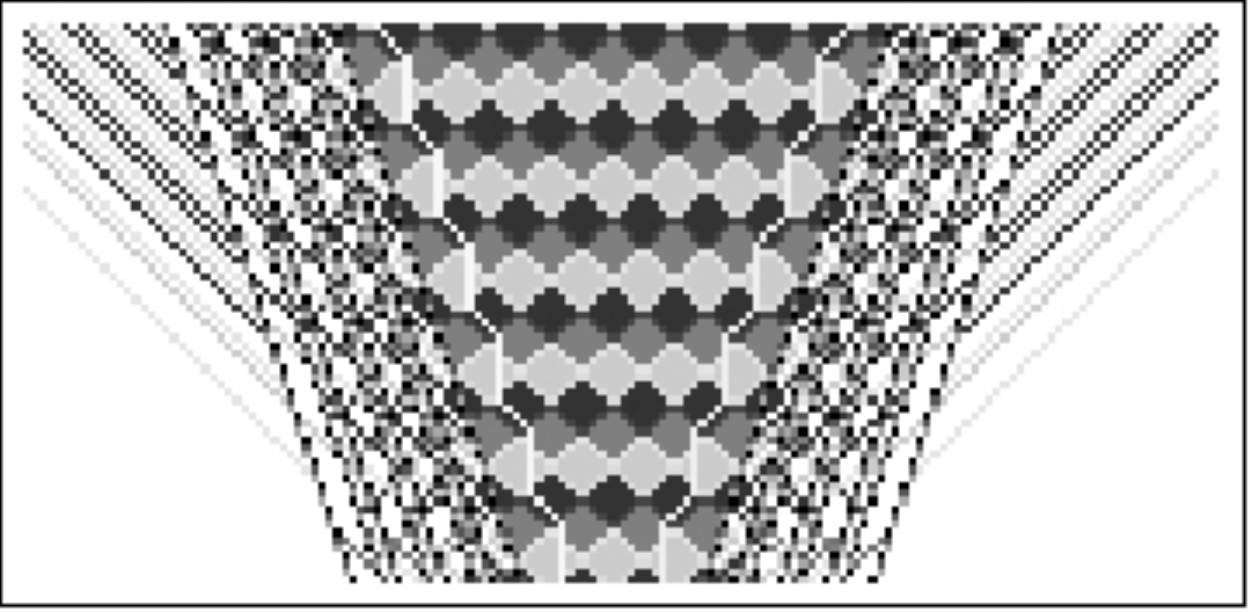}
\caption{A complete example of the QCA. The ternary background pattern is produced by the four pairs of copy bands. Two data bits are shown as two white lines propagating through the ternary background pattern. In this example no quantum gates act on the data bits. Time flows upwards.}
\label{seededqca}
\end{center}
\end{figure}

\emph{Determining the speed of computation.} When setting up our strong intrinsic universal QCA to simulate a QCA having scattering unitary $V$, the first task is to determine the size $s$ of the supercells used for the encoding. This then determines how many steps $t$ of the simulating QCA are needed to simulate one step of the simulated QCA. Refer back to Subsection \ref{subsecsim} for a formal definition of this $s$ and $t$.\\
This step is very much identical to that of Subsection \ref{weakresults}. Again the key elements in order to work out $s$ are the sizes of the quantum circuit description $C$ of the scattering unitary $V$, namely $2n$ the number of qubits upon which $C$ acts and $m$ the number of elementary universal quantum gates which make up $C$. Because each simulated cell will be encoded into the simulating QCA via just the same isometric coding $E$ (with all of the extra subsystems left in $\ket{0}$), we have $s=4nm+2+2n$ and $t=(3/2)s$ once more.\\ 
But then the number, width and the separation of the copy bands needs be worked out too. The number of copy bands is easily found: since two copy bands are necessary to encode one gate, there are $4mn+2$ ($4mn$ for program signals and $2$ for change colour signals) copy bands in total. These $4mn+2$ copy bands are placed symmetrically at the left and right side of the data, with each side containing $2mn+1$ copy bands.\\
The rate at which the ternary background signals are fired is equal to the distance traveled by the bouncer signal from the inner wall to the outer wall (refer to Figure~\ref{copyband} for a visualization). We want this rate to be $s/2$, because there are $s/2$ ternary background signals to be fired on each side. Let us call $d$ the initial value of the width of each band (the distance between the inner wall and the outer wall, excluding walls). Clearly this is what determines the rate at which the ternary background signals are fired, yet this rate is not just $d$, because in traveling this distance, the bouncer signal has to `catch up' with the outer wall which is moving at speed one third, and hence 
\begin{align*}
\frac{s}{2}&=d+\frac{1}{3}d+\frac{1}{9}d+\frac{1}{27}d+\ldots=\frac{1}{1-(1/3)}d=\frac{3}{2}d\\
\textrm{so}\quad d&=s/3=\frac{1}{3}(4mn+2+2n)
\end{align*}
Here the fact that $\frac{1}{3}s$ may not be an integer arises from the flexibility of the initial configuration, namely the flexibility in choosing initial values of the \textbf{counter} subsystem. As a result, it is possible to find two values of $d$, one equals to $\lfloor\frac{1}{3}s\rfloor$ and the other one equals to $\lceil\frac{1}{3}s\rceil$, by adjusting the initial values of the \textbf{counter} subsystem in the inner and outer walls, that produce the same result (i.e. correspond to the same value of $s$). Note however, the choice of $d$ and the initial values of the \textbf{counter} subsystem is not arbitrary, because the current neighborhood scheme of the QCA may render the walls moving in the wrong direction. If all the \textbf{counter}s start in the same value, we can fix the value of $d$ by round $\frac{1}{3}s$ to the closest integer.
%We need to have $\frac{s}{2}=\lfloor\frac{(d-2)}{3}\rfloor+(d-2)$ , where $(d-2)$ is the distance between the walls, and $\lfloor\frac{(d-2)}{3}\rfloor$ is the extra distance the bouncer signal has to travel to `catch up' with the shifted wall.  Here we are only using the `returning' part of the bouncer signal (i.e. the part where the bouncer signal travels from the inner wall to the outer wall) to measure $s$, because this part is the distance between two successively fired program signals, as it is perpendicular to both of them (refer to Figure~\ref{copyband} for a visualization). So we have $d=\frac{3}{8}s+2$. $s$ is also the number of program signals we must generate to fully weave the ternary background.\\
%Note that if we set the counters at the inner wall and outer wall at the same value when they start, both walls shift at the same time, and so the width of the wall is always $d$. However we always want the bouncer signal to bounce upon a the wall at a moment where the wall is not shifting, an the easiest way to arrange for that  is that sometimes the counters of the inner wall and the outer wall do not start at the same value. Hence the actual width of the copy band may `oscillate' between $d-1$ and $d+1$. With need to keep this in mind in order to fire program signals at appropriate intervals. \\
The separation between different copy bands just needs to be done by leaving one quiescent cell in between them, so that the proximity of each others wall do not interfere with their functioning.\\
Now by multiplying the widths and separations of the bands together with the number of bands, we obtain the size of the $r$ parameter of the strong isometric coding, as was formally defined Subsection \ref{subsecsim}. In our scheme we have: \[r=(\frac{1}{3}s+2+1)*(2mn+1)=(\frac{1}{3}(4mn+2+2n)+3)*(2nm+1)\]\\
Notice a nice feature of our construction: our copy bands move apart at speed $1/3$, but the expansion rate of the simulated QCA is at most $s/2t$, which is also $1/3$. Hence the ternary background pattern, which is being constantly `weaved' by our copy bands, grows fast enough so that the data it holds never `leak' out of it. This point is illustrated in Figure \ref{seededqca}.

\emph{Prepare the initial bouncer and history value.}   The next step would be to figure out the initial \textbf{bouncer} value, and the corresponding \textbf{history} value. The best way to do this is to use a fixed \textbf{bouncer} value for all copy bands, and use the technique of subsection~\ref{modesync} to trace the `hypothetical' change colour signals, and calculate the appropriate \textbf{history} value for each band by hand.

A complete example of the strong intrinsic universal QCA is shown in Fig.~\ref{seededqca}. It has four program bits, hence four copy bands at each side. The bands are numbered 1 to 8 from left to right, and their initial configurations are shown in Table~\ref{initconf}. This figure does not show the actual interaction between gates and data, so data just `flow through' the ternary background pattern.

\begin{table}[!hbp]
\begin{scriptsize}
	\centering
		\begin{tabular}{|c|c|p{10cm}|}
			\hline
				Band&Configuration \\
				\hline
				\hline
				1&$|30130000\rangle$, $|00001300\rangle$, $|00000000\rangle$, $|00000000\rangle$, $|30200001\rangle$\\
				\hline
				2&$|30130000\rangle$, $|00001300\rangle$, $|00000000\rangle$, $|00000000\rangle$, $|30200421\rangle$\\
				\hline
				3&$|30130000\rangle$, $|00001300\rangle$, $|00000000\rangle$, $|00000000\rangle$, $|30200330\rangle$\\
				\hline
				4&$|30110420\rangle$, $|00003400\rangle$, $|00000000\rangle$, $|00000000\rangle$, $|30200322\rangle$\\
				\hline
				5&$|30200322\rangle$, $|00000000\rangle$, $|00000000\rangle$, $|00003400\rangle$, $|30110410\rangle$\\
				\hline
				6&$|30200310\rangle$, $|00000000\rangle$, $|00000000\rangle$, $|00001300\rangle$, $|30130000\rangle$\\
				\hline
				7&$|30200411\rangle$, $|00000000\rangle$, $|00000000\rangle$, $|00001300\rangle$, $|30130000\rangle$\\
				\hline
				8&$|30200001\rangle$, $|00000000\rangle$, $|00000000\rangle$, $|00001300\rangle$, $|30130000\rangle$\\
				\hline
		\end{tabular}
	\caption{Initial configurations of each copy band. The subsystems in each cell is ordered by \textbf{$|counter|data|wall|copy|bouncer|program|mode|history\rangle$}}
	\label{initconf}
	\end{scriptsize}
\end{table}

\subsection{Formal results}
The strong intrinsic universal QCA shares many formal properties with the weak one. The way that universal quantum gates are constructed and addressed, as described in subsection~\ref{weakresults}, can be applied directly to the strong intrinsic universal QCA. The additional structures, namely the copy bands, correspond to the $L$ and $R$ isometries in Definition~\ref{strongisomcode}. With the construction details described in this section, we have shown the following theorem:
\begin{Th}
There exists $G'$ a $U$-defined QCA which is strongly intrinsically universal QCA in the following sense. Let $G$ be $V$-defined QCA such that $V$ can be expressed as a quantum circuit $C$ made of $m$ gates acting upon $2n$ qubits. Then $G'$ is able to strongly intrinsically simulate $G$ with space expansion factor $s=4nm+2+2n$, time expansion factor $t=(3/2)s$. Moreover if the initial configuration we want to simulate is of size $x$, then the initial configuration of the simulating configuration is of size $sx+2r$ with $r=(\frac{1}{3}(4mn+2+2n)+3)*(2nm+1)$.
\end{Th}
\textbf{Proof.} The $E$ encoding is the same as the one in Subsection \ref{weakresults}, and the $L$ and $R$ are the constructions of the let and right `copy bands' described in this section. Their different sizes have been worked out previously.\\
\hfill $\Box$

\section{Conclusion}\label{conclusion}

\noindent \emph{Main claim and future work.} We have formalized the notion of a QCA capable of simulating all others with linear overhead, exactly if the scattering unitaries they are made of decompose into a circuit of elementary quantum gates, and approximately otherwise. We have constructed such an intrinsic universal QCA, which turns out to be a Partitioned QCA (Figure \ref{structure}) of cell-dimension $36$ and whose scattering unitary we have given explicitly (Figure \ref{Uascircuit} and Subsection \ref{theqca}). If we insist that the initial configuration of the simulating QCA be finite (and not just periodic) we get to a stronger notion of universality, which was also formalized here. We have also constructed such a strong intrinsic universal QCA, which turns out to be a Partitioned QCA (Figure \ref{structure}) of cell-dimension $15 552$ and whose scattering unitary we have given explicitly (Figures \ref{layers} to \ref{program_layer} and Subsection \ref{strongfitting}).\\
Clearly the main challenge we now face is to find an intrinsically universal QCA in $n>1$ dimensions. The construction described in this paper is unlikely to be useful then, as it seems to blow up in complexity. But surprising as it may seem, in the classical reversible cellular automata literature intrinsically universal $n$-dimensional reversible cellular automata turn out have appeared before the intrinsically one-dimensional reversible cellular automata \cite{Ollinger2}. The common belief in the classical CA community seems to be that the $n$-dimensional constructions are actually simpler, due to the ability to use up the more than one-dimensional space to draw up circuits and sort out wire cross-overs. Intuitively at least, in $n$-dimensions intrinsic universality is very much the same as circuit universality. This is the direction which we plan to take in the close future.

\section*{Acknowledgements}
P.J.A would like to thank Torsten Franz, Miguel Lezama and Philippe Jorrand for their support.

\end{document}